\documentclass{aa}  

\usepackage{graphicx}
\usepackage{amsmath}
\usepackage{txfonts}
\usepackage{hyperref}
\usepackage{natbib}
\bibpunct{(}{)}{;}{a}{}{,} 

\begin{document}

   \title{USco1621 B and USco1556 B: Two wide companions at the\\ deuterium-burning mass limit in Upper Scorpius}


   \author{Patricia Chinchilla
          \inst{1,2}
          \and
          V\'ictor J. S. B\'ejar\inst{1,2}
          \and
          Nicolas Lodieu\inst{1,2}
          \and
          Bartosz Gauza\inst{3,4}
          \and
          Maria Rosa Zapatero Osorio\inst{5}
          \and
          Rafael Rebolo\inst{1,2,6}
          \and
          Antonio P\'erez Garrido\inst{7}
          \and
          Carlos Alvarez\inst{8}
          \and
          Elena Manjavacas\inst{8}
          }

   \institute{Instituto de Astrof\'isica de Canarias (IAC), Calle V\'ia L\'actea s/n, 38200 La Laguna, Tenerife, Spain.
              \\
              \email{pcg@iac.es}
         \and
             Departamento de Astrof\'isica, Universidad de La Laguna (ULL), 38205 La Laguna, Tenerife, Spain.
             \and
             Departamento de Astronom\'ia, Universidad de Chile, Camino El Observatorio 1515, Las Condes, Santiago, Chile.
             \and
             Janusz Gil Institute of Astronomy, University of Zielona G\'ora, Lubuska 2, 65-265 Zielona G\'ora, Poland.
             \and
             Centro de Astrobiolog\'ia (CSIC-INTA), Ctra. de Ajalvir km 4, 28850 Torrej\'on de Ardoz, Madrid, Spain.
             \and
             Consejo Superior de Investigaciones Cient\'ificas (CSIC), Spain.
             \and
             Departamento de F\'isica Aplicada, Universidad Polit\'ecnica de Cartagena, Campus Muralla del Mar, 30202 Cartagena, Murcia, Spain.
             \and
             W. M. Keck Observatory, 65-1120 Mamalahoa Highway, Kamuela, HI 96743, USA.
             }

   \date{Received 18 June 2019 / Accepted 29 November 2019}

 
  \abstract
   {}
   {Our objective is to identify analogues of gas giant planets, but located as companions at wide separations of very young stars. The main purpose is to characterise the binarity frequency and the properties of these substellar objects, and to elucidate their early evolutionary stages. 
   }
   {To identify these objects, we cross correlated the Visible and Infrared Survey Telescope for Astronomy (VISTA) Hemisphere Survey (VHS) and the United Kingdom Infrared Telescope Infrared Deep Sky Survey Galactic Clusters Survey (UKIDSS GCS) catalogues to search for common proper motion companions to 1195 already known members of Upper Scorpius (USco; age $\sim$5--10 Myr, distance $\sim$145 pc). We present the discovery and spectroscopic characterisation of two very wide substellar companions of two early-M stars in Upper Scorpius: USco1621\,B and USco1556\,B. We obtained optical and near-infrared low-resolution spectroscopy of the candidates 
   to characterise their spectral energy distribution and confirm their youth and membership to the association. We also acquired adaptive optics images of the primaries and secondaries to search for signs of binarity and close companions.}
   {By comparison with field dwarfs and other young members of USco, we determined a 
   spectral type of M8.5 in the optical for both companions, along with L0 and L0.5 in the near-infrared for USco1621\,B and USco1556\,B, respectively. The spectra of the two companions show evident markers of youth, such as weak alkaline Na~I and K~I lines, along with the triangular shape of the $H$-band. The  comparison with theoretical evolutionary models gives estimated masses of 0.015$\pm$0.002 and 0.014$\pm$0.002 M$_{\odot}$ , with temperatures of 2270$\pm$90 and 2240$\pm$100 K, respectively. The physical separations between the components of both systems are 2880$\pm$20 and 3500$\pm$40 AU for USco1621 and USco1556 systems, respectively. We did not find any additional close companion in the adaptive optics images. The probability that the two secondaries are physically bound to their respective primaries, and not chance alignments of USco members, is 86\%, and the probability that none of them are physically related is 1.0\%.}
   {}

\titlerunning{USco1621 B and USco1556 B. Two wide companions at the deuterium-burning limit in USco.}
   \maketitle
%

\section{Introduction}

\begin{table*}
\tiny
\caption{Known wide substellar companions with projected separations above 1000 AU}             
\label{table:widecompilation}      
\centering          
\begin{tabular}{llcccccc} 

 \hline
\noalign{\smallskip}                    
 & & & & Age & Projected & Companion &  \\
 Name & Short Name & RA & DEC & (Gyr) \tablefootmark{a} & Sep. (AU) \tablefootmark{a} & Mass ($M_{\mathrm{Jup}}$) \tablefootmark{a}& Ref.\tablefootmark{c} \\
\hline
\noalign{\smallskip} 

  2MASS J160251.16$-$240150.2 & USco 1602$-$2401 B & 16:02:51.17 & $-$24:01:50.45 & 0.005--0.010 & 1000 & 19--67 & 1 \\
 SR12 C & SR12 C & 16:27:19.51 & $-$24:41:40.1 & 0.0003--0.010 & 1083 & 6--21 & 2 \\
 ULAS J130041+122114  & Ross 458 c & 13:00:41.73 & $+$12:21:14.7 & $<$1 & 1168 & 5--14 & 3 \\
 ULAS J150457.65+053800.8 & HIP 73786 B & 15:04:57.66 & $+$05:38:00.8 & $>$1.6 & 1260 & ... \tablefootmark{b} & 4, 5 \\
2M13480290$-$1344071 &  2M1348$-$1344B & 13:48:02.90 & $-$13:44:07.1 & 4--10 & 1400 & 31--79 & 6, 7 \\
 2MASS J06462756$+$7935045 & HD 46588 B & 06:46:27.56 & $+$79:35:04.5 & 1.3--4.3 & 1420 & 47--75 & 8 \\
 $\varepsilon$ Indi Ba & $\varepsilon$ Indi Ba & 22:04:10.52 & $-$56:46:57.7 & 1.3 & 1459 & 37--57 & 9, 10 \\
 $\varepsilon$ Indi Bb & $\varepsilon$ Indi Bb & 22:04:10.52 & $-$56:46:57.7 & 1.3 & 1459 & 21--35 & 9, 10 \\
 2MASS J1457150$-$212148 & Gl 570D & 14.57:15.04 & $-$21:21:49.82 & 2--10 & 1525 & 30--70 & 11 \\
 HIP 38939 B & HIP 38939 B & 07:58:01.61 & $-$25:39:01.4 & 0.3--2.5 & 1630 & 18--58 & 12 \\
 2MASS J04414489+2301513 Ba & 2M0441$+$2301 Ba & 04:41:45.65 & $+$23:01:58.0 & 0.001--0.003 & 1800 & 16--22 & 13, 14 \\
2MASS J04414489+2301513 Bb  & 2M0441$+$2301 Bb & 04:41:45.65 & $+$23:01:58.0 & 0.001--0.003 & 1800 & 8--12 & 13, 14 \\
2MASS J02495436$-$0558015  & 2MASSJ0249$-$0557 c & 02:49:54.36 & $-$05:58:01.5 & 0.016--0.028 & 1950 & 10.6--12.9 & 15 \\
Gl 417B & Gl 417B & 11:12:25.7 & $+$35:48:13 & 0.63--0.9 & 1970 & 50--56 & 16, 17, 18 \\
 Gl 417C & Gl 417C & 11:12:25.7 & $+$35:48:13 & 0.63--0.9 & 1970 & 45--52 & 16, 17, 18 \\
GU Psc b  & GU Psc b & 01:12:36.48 & $+$17:04:31.8 & 0.07--0.13 & 2000 & 9--13 & 19 \\
 WD 0806-661 B & WD 0806-661 B & 08:07:14.68 & $-$66:18:48.7 & 1.5--2.5 & 2500 & 6--9 & 20, 21 \\
 SDSS J224953.47+004404.6A & SDSS J2249$+$0044A & 22:49:53.47 & $+$00:44:04.6 & 0.012--0.790 & 2600 & 11--73 & 22  \\
 SDSS J224953.47+004404.6B & SDSS J2249$+$0044B & 22:49:53.47 & $+$00:44:04.6 & 0.012--0.790 & 2600 & 9--68 & 22 \\
 WISEP J142320.86+011638.1 & BD +01$^{\circ}$ 2920B & 14:23:20.86 & $+$01:16:38.1 & $>$2.3 & 2630 & 20--50 & 23 \\
SDSS J175805.46+463311.9  & SDSS J1758$+$4633 & 17:58:05.46 & $+$46:33:11.9 & 0.5--1.5 & 2685 & 21--37 & 24 \\
 2MASS J16212830$-$2529558 & USco1621 B & 16:21:28.31 & $-$25:29:56.1 & 0.005--0.010 & 2900 & 14--18 & 25 \\
 WISE J200520.38+542433.9 & Wolf 1130C & 20:05:20.38 & $+$54:24:33.9 & $>$2 & 3150 & $>$52 & 26 \\
 HIP 77900B & HIP 77900B & 15:54:30.47 & $-$27:19:57.51 & 0.005--0.010 & 3200 & 15--27 & 1 \\
2MASS J15562344$-$2541056  & USco1556 B & 15:56:23.43 & $-$25:41:05.7 & 0.005--0.010 & 3500 & 12--17 & 25 \\
 2MASSW J1523226+301456 & Gl 584C & 15:23:22.6 & $+$30:14:56 & 1.0--2.5 & 3600 & 47--79 & 16 \\
 SDSS J213154.43$-$011939.3 & SDSS 2131$-$0119 & 21:31:54.43 & $-$01:19:39.3 & $>$1 &  3800 & 52--73 & 27 \\
 WISE J111838.70+312537.9 & WISE 1118+31 & 11:18:38.70 & $+$31:25:37.9 & 2--8 & 4100 & 14--38 & 28 \\
 VVV J151721.49$-$585131.5 & $\beta$ Cir B & 15:17:21.60 & $-$58:51:30.0 & 0.37--0.50 & 6656 & 51--66 & 29 \\
 2MASS J21265040$-$8140293 & 2MASS J2126$-$8140 & 21:26:50.40 & $-$81:40:29.3 & 0.010--0.045 & 6900 & 11.6--15 & 30 \\
HIP 70849B  & HIP 70849B & 14:28:42.32 & $-$05:10:20.9 & 1--5 & 9000 & ... \tablefootmark{b} & 31 \\
 ULAS J133943.79+010436.4 & HD 118865B & 13:39:43.79 & $+$01:04:36.4 & 1.5--4.9 & 9200 & 42--68 & 32 \\
 ULAS J145935.25$+$085751.2 & ULAS J1459$+$0857 & 14:59:35.25 & $+$08:57:51.2 & $>$4.8 & $\sim$20\,000 & 63--79 & 33 \\
\hline
                 
\end{tabular}

\tablefoot{
\tablefoottext{a}{Data compiled from the literature.}
\tablefoottext{b}{Classified as a T dwarf.}
\tablefoottext{c}{ References: (1) \citet{Aller2013}; (2) \citet{Kuzuhara2011}; (3) \citet{Goldman2010}; (4) \citet{Scholz2010}; (5) \citet{Murray2011}; (6) \citet{Muzic2012}; (7) \citet{Deacon2012}; (8) \citet{Loutrel2011}; (9) \citet{Scholz2003}; (10) \citet{McCaughrean2004}; (11) \citet{Burgasser2000}; (12) \citet{Deacon2012_2}; (13) \citet{Todorov2010}; (14) \citet {Bowler2015}; (15) \citet{Dupuy2018}; (16) \citet{Kirkpatrick2001}; (17) \citet{Bouy2003}; (18) \citet{Dupuy2014}; (19) \citet{Naud2014}; (20) \citet{Luhman2011}; (21) \citet {Luhman2012_2}; (22) \citet{Allers2010}; (23) \citet{Pinfield2012}; (24) \citet{Faherty2010}; (25) This work; (26) \citet{Mace2013}; (27) \citet{Gauza2019}; (28) \citet{Wright2013}; (29) \citet{Smith2015}; (30) \citet{Deacon2016}; (31) \citet{Lodieu2014}; (32) \citet{Burningham2013}; (33) \citet{Day-Jones2011}}
} 

\end{table*}

Multiplicity is an important outcome of the formation processes giving rise to stars and planetary systems and is key to understand the stellar and substellar physics. Binarity is a common phenomenon and the latest results indicate that a large fraction of stars form as part of multiple systems \citep{Tokovinin2012, Raghavan2010} and that the binary frequency and the separation between components decrease when the mass of the primary decreases \citep[][ and references therein]{LuhmanREVIEW2012, Cortes2017}.

The substellar companions discovered at wide separations from their primaries allow for  a detailed photometric and spectroscopic characterisation to be carried out, which is extremely difficult in the case of objects found at close orbits around their primaries. The identification of substellar objects in binary systems also allows for  an inference with regard to the ages, distances, and metallicities from their brighter and easier to characterise primaries, and to determine the otherwise ambiguous luminosities, effective temperatures, and masses of the companions \citep[e.g.][]{Rebolo1998, Faherty2010, Deacon2014, Gauza2015}. Young moving groups, stellar clusters and associations with well determined ages, metallicities and distances are also helpful in the determination of the physical parameters of substellar objects.

Low-mass companions at wide orbits typically have very low gravitational binding energies, so they are not expected to survive as bound systems for a long time, especially in dense environments since they can be easily disrupted by external dynamical perturbations \citep{Kroupa1995, Close2007}.  
However, several substellar companions of all ages with orbital separations of more than a thousand AU have been found (see Table \ref{table:widecompilation} and references therein).

Dedicated studies have also been performed to investigate the multiplicity at wide separations in young moving groups and associations. For example, \citet{Kraus2011} carried out a search for wide binaries in the Taurus-Auriga star-forming region, exploring separations of 3--5000 AU, finding that only 25--32\% of the stars in the region are single, whereas the majority of them are part of a multiple system, either with stellar or substellar secondaries. \citet{Aller2013} explored the frequency of substellar companions between 15--60 M$_{\mathrm{Jup}}$ at orbital distances of 400--4000 AU in the Upper Scorpius (USco) region, finding a companionship rate of 0.6$\pm$0.3\%. \citet{Elliott2016} studied the distribution of the binary population at wide separations in the $\beta$ Pictoris moving group, finding stellar and substellar companion candidates up to physical separations of 100\,000 AU. From a sample of 49 $\beta$ Pictoris members and systems, they found 14 stellar systems ($\sim$29\%) with separations above 1000 AU, and seven stellar systems ($\sim$14\%) with separations above 10\,000 AU. They also found four substellar companions with separations $<$1000 AU.

The USco region is part of the Scorpius Centaurus (Sco-Cen) Association, the nearest OB association to the Sun. USco has an estimated average age of $\sim$ 5--10 Myr \citep{Preibisch2002, Slesnick2006, Lodieu2008, Pecaut2012, Song2012, Feiden2016, PecautMamajek2016, Rizzuto2016, Fang2017, David2019} and is located at a heliocentric distance of 146$\pm$3$\pm$6 pc, where the second term in the uncertainty refers to systematic errors \citep{Galli2018}. The mean proper motion of USco members is ($\mu_{\alpha}$ cos $\delta$, $\mu_{\delta}$) = ($-$10, $-$23) $\pm$ 6 mas yr$^{-1}$ \citep{Fang2017}. Its youth and close distance make this association an ideal place for the study of substellar objects, as these objects cool down with time and are still relatively bright at such young ages.
A number of wide low-mass substellar companions with masses below or close to the deuterium-burning mass limit \citep[$\sim$13 M$_{\mathrm{Jup}}$ for solar metallicities, ][]{Burrows1997, Burrows2001} have been found in the USco region, such as GSC 06214-00210 \citep{Ireland2011}, and 1RXS1609b \citep{Lafreniere2008}. These companions orbit young K-type stars at physical separations below 500 AU. A less massive system, UScoCTIO 108 AB \citep{Bejar2008}, formed by a brown dwarf and a planetary-mass companion at the deuterium-burning mass limit was also found at the wider separation of 670\,AU in this region. 

We have performed a search for wide binaries in USco, at separations between $\sim$400--9000 AU,  to assess the frequency of wide companions to young stars and compare with the companionship rate for different ages. Our main goal is to provide observational constraints to evaluate the different wide-binary formation models and test general scenarios of substellar formation. In this work, we present the identification of 2MASS J16212830$-$2529558 (hereafter, USco1621\,B) and 2MASS J15562344$-$2541056 (hereafter, USco1556\,B) as two wide substellar companions to the previously-known low-mass stars of USco 2MASS J16212953$-$2529431 (hereafter, USco1621~A) and  2MASS J15562491$-$2541202 (hereafter, USco1556~A) \citep{Rizzuto2015}. In Section 2, we present the search method. Section 3 describes the follow-up observations of the systems, data reduction, and immediate results derived from these observations. In Section 4, we analyse the physical properties of the secondaries. Section 5 is dedicated to the discussion of the companionship of these pairs and the analysis of their formation and evolution. Finally, Section 6 corresponds to the summary and final remarks.


\section{Two new young, low-mass systems}

   \begin{figure}
   \centering
   \includegraphics[width=9cm]{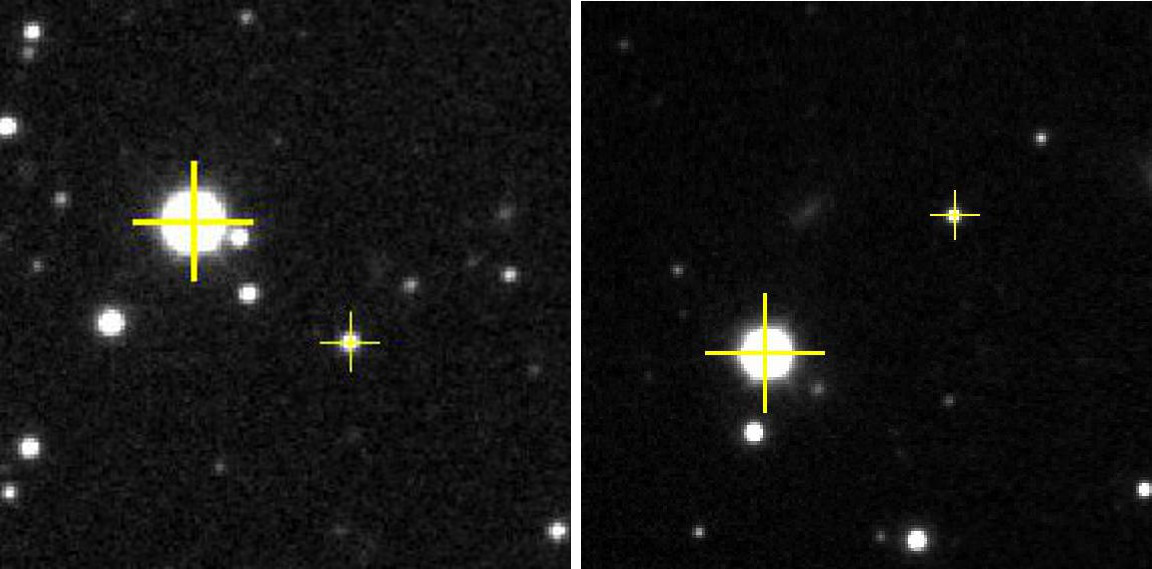}
   \caption{VHS \textit{J}-band images of USco1621~AB (left) and USco1556~AB (right). The primaries and secondaries are marked with crosses. The field of view is  1\arcmin$\times$1\arcmin and the orientation is north up and east to the left.}
              \label{fig:Imagendirecta}%
    \end{figure}

\subsection{Astrometric and photometric search}

   \begin{figure}
   \centering
   \resizebox{\hsize}{!}{\includegraphics{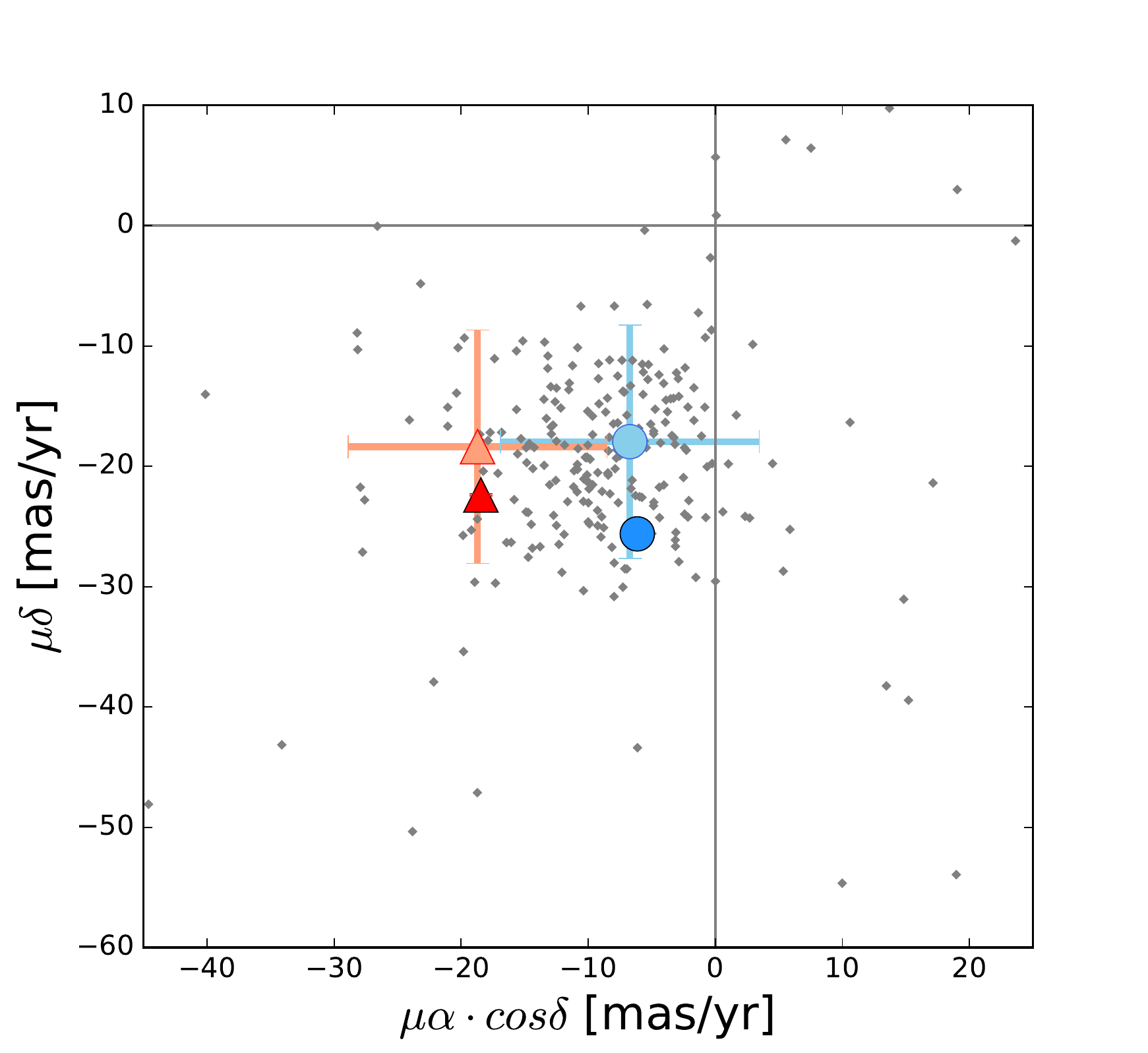}}
   \caption{Proper motion diagram of USco1621~A and B (blue circles) and USco1556~A and B (red triangles). The proper motions of USco known members from our VHS vs UKIDSS GCS cross-correlation are plotted with grey dots. The proper motions of the secondaries are calculated from the VHS and UKIDSS GCS astrometric data and plotted in lighter colours, while the proper motions of the primaries are obtained from $Gaia$ DR2 and plotted in darker colours. The error bars of the $Gaia$ DR2 measurements are smaller than the size of the symbols.}
              \label{fig:PM}%
    \end{figure}

We have performed a search for companions orbiting at wide separations from a list of 1195 known member candidates of the USco association compiled from the literature \citep{Walter1994, Preibisch1998, Ardila2000, Martin2004, Lodieu2006, Slesnick2006, Lodieu2007, Findeisen2010, Dawson2011, Lodieu2011, Luhman2012, Dawson2013, Lodieu2013, Rizzuto2015, Best2017}. For this purpose, we have made use of proprietary data from the Visible and Infrared Survey Telescope for Astronomy (VISTA) Hemisphere Survey \citep[VHS, ][]{McMahon2013} catalogue in combination with public data from the United Kingdom Infrared Telescope Infrared Deep Sky Survey \citep[UKIDSS, ][]{UKIDSSpaper} Galactic Clusters Survey (GCS) catalogue.

The VISTA telescope (4m diameter quasi-Ritchey-Chretien telescope) is located at ESO's Cerro Paranal Observatory (Chile), and it is equipped with a near-infrared camera, VISTA InfraRed CAMera (VIRCAM), which provides a field of view of 1.65 degree diameter using a mosaic of 16 detectors, adding a total of 67 million pixels with a projected size in the sky of 0.34\arcsec pix$^{-1}$ \citep{VISTApaper}. The VHS is a near-infrared survey which images the entire southern hemisphere ($\sim$ 20\,000 square degrees) in \textit{J} and \textit{Ks} bands. It also includes \textit{H} band in 5000 square degrees in the South Galactic Cap, and \textit{H} and \textit{Y} bands in the remaining areas of high Galactic latitude ($|b|>30^\circ$). It reaches limiting magnitudes down to \textit{J}=20.2 and \textit{Ks}=18.1 (Vega system), around 4 magnitudes deeper than the previous infrared surveys 2MASS \citep{2MASSpaper} and DENIS \citep{DENISpaper}. The astrometric solution of VHS is provided by the VISTA Data Flow System pipeline \citep{Irwin2004, Lewis2010} at the Cambridge Astronomical Survey Unit (CASU), which uses the 2MASS Point Source Catalog astrometry as a reference, giving a relative accuracy that is better than 0.1\arcsec.\footnote {for further details, see \url{http://casu.ast.cam.ac.uk/surveys-projects/vista/technical}.}

The UKIDSS GCS is a survey performed by the 3.8m United Kingdom Infrared Telescope (UKIRT) which is located on the summit of Maunakea in Hawai'i. The UKIRT telescope is equipped with the UKIRT Wide Field Camera \citep[WFCAM;][]{WFCAMpaper}, which provides a field of view of 0.9 degree diameter, using four 2048$\times$2048 HgCdTe Rockwell Hawaii-2 infrared detectors with a pixel size of 18 microns and a plate scale of 0.40\arcsec pix$^{-1}$. The UKIDSS GCS covers an area of $\sim$ 1600 square degrees around several Galactic open clusters and star-forming associations,  in the \textit{ZYJHK} filters, up to limiting magnitudes of \textit{J} $\sim$ 19.1, \textit{K} $\sim$ 18.4. One of the regions covered by GCS is the USco region. The astrometric calibration is performed by CASU in a similar way to VISTA, giving an rms accuracy per source better than 0.1\arcsec \citep{Hambly2008, Hodgkin2009}.

Our compiled catalog contains 1306 individual members of USco, grouped in 1286 systems. 1195 of these 1286 systems are in the region of USco covered by both VHS and GCS catalogues.
Companion candidates were identified as follows: we singled out all the detected sources in a circular area of 60\arcsec radius (corresponding to $\sim$9000 AU at the USco distance) around each known USco member. We adopted this value as a compromise between the expected number of contaminants and the number of true companions. We cross-correlated the VHS and UKIDSS GCS catalogues using TOPCAT \citep{TOPCATpaper} but limiting the cross-correlation exercise to the aforementioned circular areas.
 The cross-match radius for the individual objects was 1\arcsec. This relatively large radius takes into account the astrometric uncertainties of the catalogues and the expected motion of USco members over the interval of seven years, which is the mean time baseline between the two surveys. In the 60\arcsec radius area around each USco member, we calculated the proper motions of all the cross-matched objects using VHS and GCS astrometry, and selected all the sources that have proper motions compatible with the mean USco motion ($-$7.46, $-$19.82, according to our measurement) with a maximum deviation of 23.5 mas yr$^{-1}$ in the total proper motion. This astrometric threshold is defined by the quadratic sum of the dispersion of the proper motions in both axes of non-moving background objects, which is dominated by faint stars and unresolved galaxies, even fainter than our discoveries. 

Besides sharing similar proper motions, we also imposed an additional condition on the astrometric candidates: they must follow the well-known USco photometric sequence in the $J$ versus $Z-J$ and $J$ versus $J-Ks$ colour-magnitude diagrams. We used the $Z$, $J$ and $Ks$ photometry of the USco known members in our compilation to determine a lower envelope limit for the candidate selection in each diagram. The photometric selection criteria were: $J_{VHS}< 4 (Z_{GCS}-J_{VHS}) + 9.5$ and $J_{VHS}< 6 (J_{VHS}-Ks_{VHS}) + 10$. For those regions not covered in the $Z$ filter by UKIDSS GCS, we used Pan-STARRS \citep{PanSTARRS2016} $z$ filter photometry to perform a similar selection. In this case, for the $J$ versus $z-J$ selection, the criterium was $J_{VHS}< 3.3 (z_{P}-J_{VHS}) + 7.7$.

Two of the faintest and reddest companion candidates found by this procedure are USco1621\,B and USco1556\,B. 
 They were not identified in the search performed by \citet{Lodieu2013}, covering the same area but using only UKIDSS GCS data. USco1621~B was missed due to the lack of $H$ filter photometry in UKIDSS GCS, and USco1556~B was missed because the proper motion in RA measured by UKIDSS GCS alone ($-19.5\pm2.97$ mas yr$^{-1}$) deviates from the mean USco motion in RA ($-8.6$ mas yr$^{-1}$) by more than three times the estimated error. USco1621\,B and USco1556\,B were, however, recognised as USco candidates by the more recent work of \citet{Luhman2018}, but these authors did not have spectroscopic information to determine their membership in the association.
Figure\,\ref{fig:Imagendirecta} shows the VHS $J$-band image of both objects and their primaries. Table\,\ref{Table:GenData} presents astrometric and photometric information for the two pairs. The angular separations are 20.78$\pm$0.02 and 24.78$\pm$0.02 arcsec for USco1621~AB and USco1556~AB, corresponding to projected orbital separations of 2910$\pm$160 and 3530$\pm$180 AU, respectively, at the distance of the primaries measured by $Gaia$ DR2 \citep{Gaiapaper, Gaiapaper2}.

 \begin{figure}
   \centering
   \includegraphics[width=9.5cm]{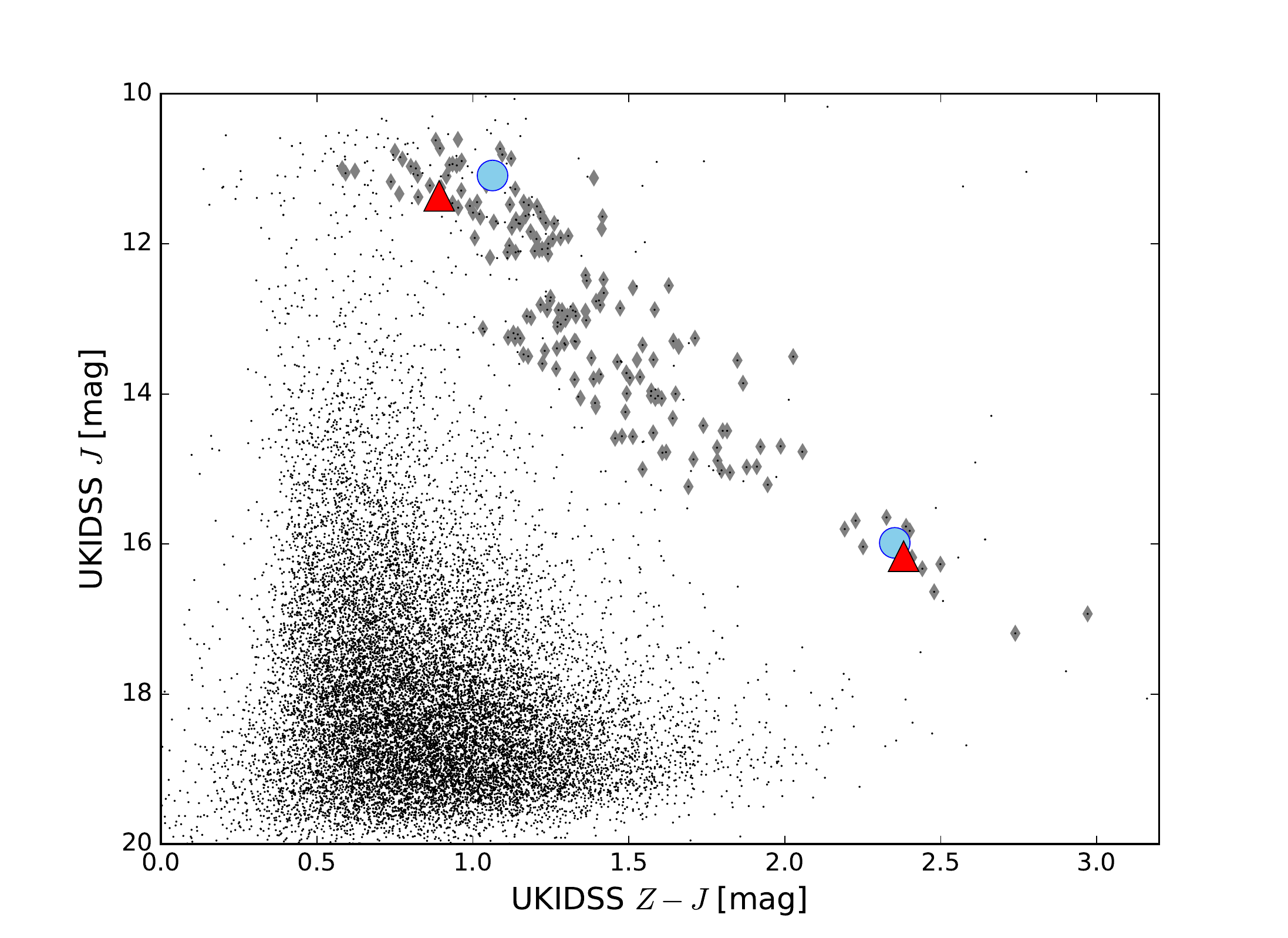}
   \caption{UKIDSS GCS \textit{J} vs \textit{Z-J} colour-magnitude diagram. Previously-known members of USco are shown as grey diamonds. Field contaminants are indicated with black dots. USco1621~A and B are marked as light blue circles, and USco1556~A and B are marked as red triangles.}
              \label{fig:colormag}%
    \end{figure}

 \begin{figure}
   \centering
   \includegraphics[width=9.5cm]{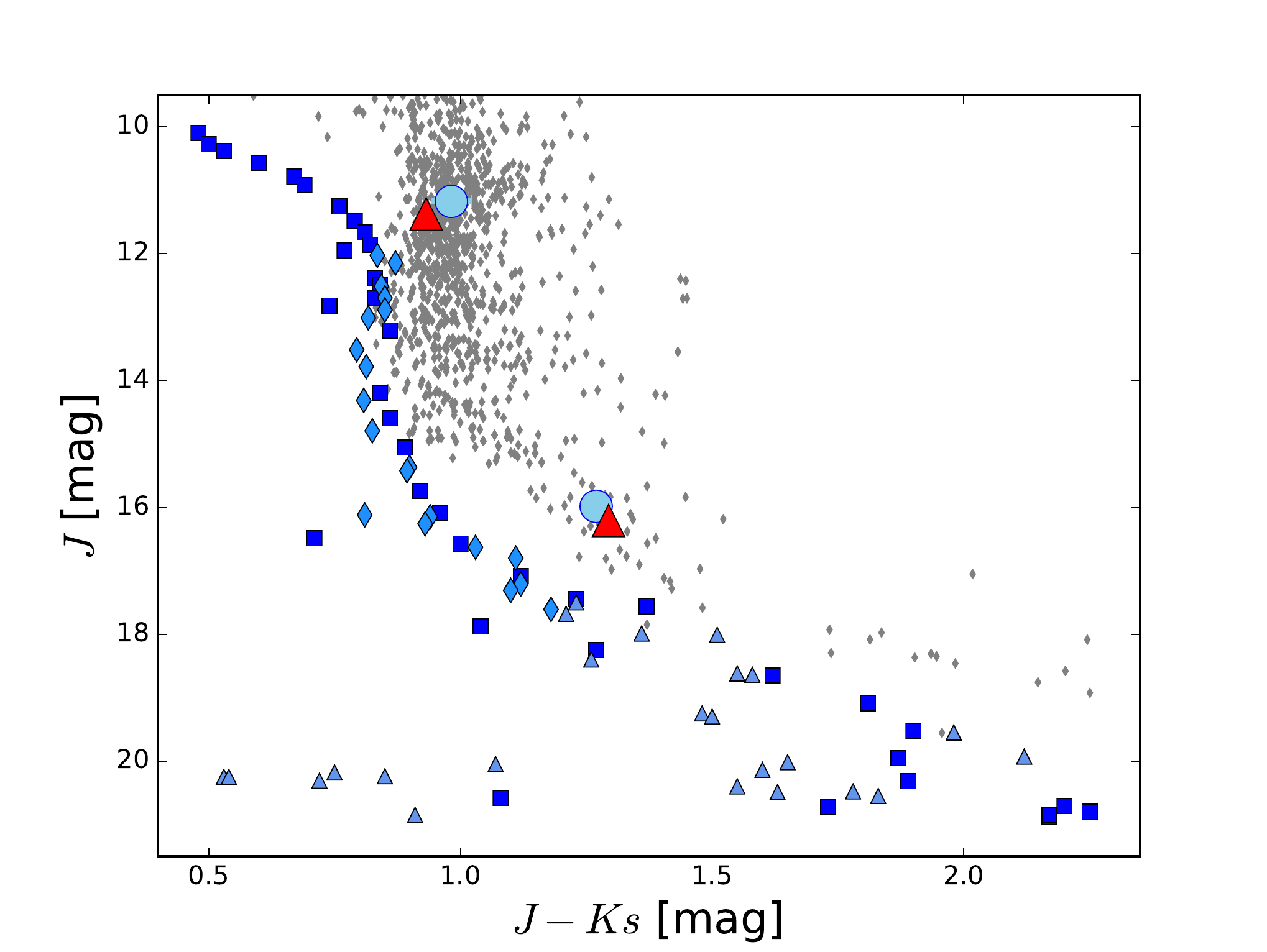}
   \caption{\textit{J} vs \textit{J-Ks} colour-magnitude diagram. USco1621~A and B are marked with light blue circles, and USco1556~A and B are marked with red triangles. The photometry of both primaries is obtained from 2MASS and the photometry of the secondaries is taken from VHS. Known members of USco are shown as grey diamonds; the photometry of objects with \textit{J}$<$14 mag is obtained from the 2MASS catalogue, while the photometry for the fainter ones is determined from VHS data. The VHS photometry has been transformed to the 2MASS photometric system through the colour equations provided by CASU. The photometric sequence of field dwarfs derived by  \citet{PecautMamajek2013} (squares) \citet{Dupuy2012} (triangles) and \citet{Lodieu2014} (diamonds), displaced to the mean distance of USco, is also shown in blue.}
              \label{fig:JvsJmK}%
    \end{figure}

%

 \begin{table*}
 \tiny
  \caption{General Data of USco1621 AB and USco1556 AB. }
  \label{Table:GenData}
  \centering
  \begin{tabular}{lcccccc}
            
   \hline
   \hline
   \noalign{\smallskip}
    \multicolumn{2}{c}{} & \multicolumn{2}{c}{USco1621 AB} & \multicolumn{1}{c}{} & \multicolumn{2}{c}{USco1556 AB} \\
    \noalign{\smallskip}
   \noalign{\smallskip}
    &&   Primary & Secondary && Primary & Secondary\\
   \noalign{\smallskip}         
   \hline
   \hline
   \noalign{\smallskip}
   Astrometry &&  & & & &\\
   \noalign{\smallskip}  
   \hline
 \noalign{\smallskip}
 
             R.A. (J2000) && 16:21:29.53  & 16:21:28.31 && 15:56:24.92  & 15:56:23.43    \\
           DEC (J2000) && $-$25:29:43.1     &   $-$25:29:56.1 && $-$25:41:20.3     &   $-$25:41:05.7  \\
            Parallax $Gaia$ (mas)  && 7.22$\pm$0.05 &   --    && 7.08$\pm$0.08 &   --    \\
            Distance (pc)   &&  138$\pm$1   & -- && 141$\pm$ 2  & --  \\
 \multicolumn{1}{l}{Separation (\arcsec)} & \multicolumn{1}{c}{} & \multicolumn{2}{c}{20.78$\pm$0.02} & \multicolumn{1}{c}{} & \multicolumn{2}{c}{24.78$\pm$0.02} \\
            \multicolumn{1}{l}{Separation (AU)} & \multicolumn{1}{c}{} & \multicolumn{2}{c}{2880 $\pm$ 20} & \multicolumn{1}{c}{} & \multicolumn{2}{c}{3500 $\pm$ 40} \\
  ($\mu_{\alpha}$cos$\delta$, $\mu_{\delta}$) $Gaia$ DR2 (mas yr$^{-1}$)&& ($-$6.14, $-$25.62) & --  && ($-$18.45, $-$22.39) &     -- \\
                     &&    $\pm$ (0.10, 0.07)  &   &&    $\pm$ (0.13, 0.08)  &     \\
($\mu_{\alpha}$cos$\delta$, $\mu_{\delta}$) VHS--UKIDSS (mas yr$^{-1}$)&& -- & ($-$7, $-$18) $\pm$ (10, 10) && -- & ($-$19, $-$18) $\pm$ (10, 10)  \\
\noalign{\smallskip}
            \hline
 \noalign{\smallskip}
              Spectroscopy && & && & \\
        \noalign{\smallskip}
            \hline
            \noalign{\smallskip}
                  Spectral type  &&  M2.5\tablefootmark{a}   & M8.5$\pm$0.5 (OPT), L0$\pm$0.5(NIR)  &&  M3.0\tablefootmark{a}   & M8.5$\pm$0.5 (OPT), L0.5$\pm$0.5 (NIR) \\
                  Li 6708 EW ($\AA$) && 0.29$\pm$0.02 \tablefootmark{a} & -- && 0.27$\pm$0.02\tablefootmark{a} & -- \\
                  H$_{\alpha}$ EW ($\AA$) && $-$4.29$\pm$0.06 \tablefootmark{a}  &  -- && $-$5.67$\pm$0.04 \tablefootmark{a} & -- \\
  \noalign{\smallskip}
            \hline
             \noalign{\smallskip}
              Photometry && & && & \\
        \noalign{\smallskip}
            \hline
             \noalign{\smallskip}

  $Gaia$ \textit{G}&&   14.1769$\pm$0.0008   &  21.11$\pm$0.03 &&   14.310$\pm$0.003   &  --  \\
         \noalign{\smallskip}

Pan-STARRS \textit{g}&&   16.117$\pm$0.003   &  --  &&   16.15$\pm$0.02   &  --  \\
                Pan-STARRS \textit{r}&&   14.706$\pm$0.004   &  -- &&   14.865$\pm$0.013   &  --  \\
                Pan-STARRS  \textit{i}&&  --    &  20.53$\pm$0.06 &&  13.68$\pm$0.03   &  21.05$\pm$0.12   \\
                Pan-STARRS  \textit{z}&&  --    &  19.11$\pm$0.05 &&  12.986$\pm$0.005    &  19.39$\pm$0.06  \\
                Pan-STARRS  \textit{y}&&  12.519$\pm$0.001    &  18.02$\pm$0.03 &&  12.670$\pm$0.002    &  18.14$\pm$0.04  \\
                \noalign{\smallskip}    

  DENIS \textit{I} &&  12.77$\pm$0.02    &  --  && -- & -- \\
                 DENIS \textit{J}&&   11.11$\pm$0.07   &  15.9$\pm$0.2  && -- & -- \\
                  \noalign{\smallskip}
GCS  \textit{Z}&&  12.1521$\pm$0.0009    &  18.34$\pm$0.04 &&  12.2534$\pm$0.0010    &  18.54$\pm$0.04  \\
                  GCS  \textit{Y}&&  11.7760$\pm$0.0008    &  17.03$\pm$0.02 &&  11.9184$\pm$0.0008    &  17.22$\pm$0.02  \\
                  GCS  \textit{J}&&  11.0888$\pm$0.0005    &  15.985$\pm$0.010 &&  11.3606$\pm$0.0008    &  16.163$\pm$0.012  \\
                  GCS  \textit{H}&& -- & -- &&  11.1524$\pm$0.0007    &  15.489$\pm$0.013  \\
                  GCS \textit{K}&&   10.5627$\pm$0.0004   &  14.636$\pm$0.006 &&   10.4900$\pm$0.0004   &  14.812$\pm$0.007  \\

                  \noalign{\smallskip}  

         VHS \textit{J} &&  --   &  15.942$\pm$0.010 &&  --   &  16.166$\pm$0.010 \\
                 VHS \textit{Ks} &&  --   &  14.704$\pm$0.011 &&  --   &  14.904$\pm$0.015  \\
                 \noalign{\smallskip}
   2MASS \textit{J}&&  11.18$\pm$0.02    &  16.17$\pm$0.08  &&  11.37$\pm$0.02    &  16.11$\pm$0.08 \\
  2MASS \textit{H}&& 10.43$\pm$0.02 & 15.34$\pm$0.10 &&  10.69$\pm$0.02    &  15.39$\pm$0.09  \\
                  2MASS \textit{Ks}&&   10.19$\pm$0.02   &  14.67$\pm$0.09 &&   10.44$\pm$0.02   &  14.85$\pm$0.12   \\
                   \noalign{\smallskip}
  WISE \textit{W1}&&  10.04$\pm$0.02    &  14.09$\pm$0.04 &&  10.32$\pm$0.02    &  14.45$\pm$0.03  \\
                  WISE \textit{W2}&&   9.90$\pm$0.02   &  13.65$\pm$0.06 &&   10.18$\pm$0.02   &  14.10$\pm$0.05  \\
                  WISE \textit{W3}&&   9.55$\pm$0.10   &  $>$11.3 &&   9.93$\pm$0.07   &  --   \\
                  \noalign{\smallskip}  
            \hline
             \noalign{\smallskip}
              Physical parameters && & && & \\
        \noalign{\smallskip}
            \hline
             \noalign{\smallskip}
                  Rot. period (days) && 2.06 \tablefootmark{b}  & -- && 4.67 \tablefootmark{b}  &  -- \\
 Log (L$_{\mathrm{bol}}$/L$_{\odot}$) &&  $-$1.02$\pm$0.03   &  $-$3.03$\pm$0.11 &&  $-$1.09$\pm$0.04   &  $-$3.07$\pm$0.11 \\
         Effective temp. (K) &&  3460$\pm$100   &  2270$\pm$90 &&  3410$\pm$100   & 2240$\pm$100 \\
         Mass (M$_{\odot}$)&&  0.36$\pm$0.08   &  0.015$\pm$0.002 &&  0.33$\pm$0.07   &  0.014$\pm$0.002 \\
        \noalign{\smallskip}

 \hline
 \hline
 
  \end{tabular}
   \tablefoot{
\tablefoottext{a}{Spectral type and Lithium and H$_{\alpha}$ equivalent widths from \citet{Rizzuto2015}}
\tablefoottext{b}{Rotation period from \citet{Rebull2018}}
} 
   \end{table*}

\subsection{USco membership}

USco1621~A and USco1556~A were discovered by \citet{Rizzuto2015} in their search for low mass members of this association. They performed a Bayesian membership selection using the photometry from 2MASS and the American Association of Variable Star Observers (AAVSO) Photometric All-Sky Survey \citep[APASS;][]{APASSpaper}, and The fourth United States Naval Observatory (USNO) CCD Astrograph Catalog \citep[UCAC4;][]{UCAC4paper} proper motions  \citep{Rizzuto2011}. 
They also carried out spectroscopic follow-up observations to confirm their membership. They determined a spectral type of M2.5 for USco1621~A and M3.0 for USco1556~A in the optical, and measured their H$\alpha$ and Li 6708\,$\AA$ equivalent widths (EWs), finding an EW(H$\alpha$) of $-$4.29\,$\AA$ and a EW(Li) of 0.29\,$\AA$ for the former, and an EW(H$\alpha$) of $-$5.67\,$\AA$ and a EW(Li) of 0.27\,$\AA$ for the later, which are values compatible with the age of USco.

$Gaia$ DR2  recently provided a more precise determination of the proper motions of USco1621~A and USco1556~A, in addition to parallactic distances of 138$\pm$1 and 141$\pm$2 pc, respectively. These proper motions and distances fit perfectly with those of the USco association.

 \citet{Rebull2018} measured the rotation periods of USco1621~A and USco1556~A using their photometric light curves from the Kepler space telescope K2 mission \citep{K2paper} and find rotation periods of  2.06 and 4.67 days, respectively. They assigned the highest probability of membership in USco to both primaries in their study. These values indicate a fast rotation, which is a hallmark of youth, since late-K and early-M dwarfs typically have rotation periods greater than 10 days for ages older than $\sim$\,100 Myr \citep[e.g.][]{Marcy1992, Kiraga2007, Engle2011, Rebull2018}.
 
 We have determined from our UKIDSS GCS--VHS correlation that the proper motions of USco1621\,B and USco1556\,B are: ($-$7, $-$18) $\pm$ (10, 10) and ($-$19, $-$18) $\pm$ (10, 10) mas yr$^{-1}$, respectively, which are compatible with the USco association. The time baseline of the measurements is nine and seven years, respectively. The errors are obtained from the dispersion of the GCS--VHS correlation of background objects with similar magnitudes and null motions. Figure \ref{fig:PM} shows the proper motion diagram of both companions together with $Gaia$ DR2 proper motions of the primaries, and the VHS--UKIDSS GCS proper motion of other USco members.
 
 The USco1621~AB and USco1556~AB systems have been previously imaged by several available optical and infrared surveys. There is photometry available for the primaries and the companion candidates in the 2MASS, DENIS, Pan-STARRS, $Gaia,$ and WISE catalogues \citep{2MASSpaper, DENISpaper, PanSTARRS2016, Gaiapaper, WISEpaper}. All these photometric data are presented in Table \ref{Table:GenData}. The apparent magnitudes and colours of the companion candidates obtained from these surveys are also consistent with young low-mass objects placed at the same heliocentric distance as USco, and these follow the photometric sequences of the association. Figures \ref{fig:colormag} and \ref{fig:JvsJmK} show the \textit{J} vs \textit{Z$-$J} and \textit{J} vs \textit{J$-$Ks} colour-magnitude diagrams, respectively. These figures show that both secondaries occupy overluminous locations in the colour-magnitude diagrams with respect to the field dwarfs, which is compatible with a very young age.

   \section{Follow-up observations and data reduction}

%
\begin{table*}
\tiny
\caption{Observations Log}             
\label{table:obslog}      
\centering          
\begin{tabular}{c c c c c c c c c c c c}     
\hline\hline       
Obs. Date & Telesc/Instrum & Mode & Grating/ & Wavelength & Slit & Pl. Scale & Disp. & Res. & Exp. & Airmass & Seeing \\ 
 &  & & Filter & Range [$\mu$m] &[\arcsec] & [\arcsec pix$^{-1}$] & [$\AA$ pix$^{-1}$] & Power & Time &  & [\arcsec] \\ 
\hline
\noalign{\smallskip}
 \multicolumn{10}{c}{USco1621 A}  \\
\hline
\noalign{\smallskip}
22 Mar 2019  & Keck-I/OSIRIS & AO & $K_{p}$ & 1.96--2.27 & -- & 0.01 & -- & --  & 150$\times$1.475s  & 1.5--1.6 & 1.0 \\   
   
 \hline
\noalign{\smallskip}
 \multicolumn{10}{c}{USco1621 B}  \\
\hline
\noalign{\smallskip}                    
 14 Jun 2005 & UKIRT/WFCAM  & Img & $Z, Y, J$ & -- & -- & 0.4 & -- & -- & 20s,20s,5s & 1.6 & 1.8 \\
 14 Mar 2011   & UKIRT/WFCAM  & Img & $K$ & -- & -- & 0.4 & -- & -- & 10s & 1.6 & 2.0 \\
 1 May 2014  & VISTA/VIRCAM  & Img & $J, Ks$ & -- & -- & 0.34 & -- & -- & 15s, 7.5s & 1.2 & 1.1 \\ 
10 Apr 2017    & NTT/SofI & Spec & GR & 1.53--2.52 & 1 & 0.29 & 10.22 & 600 & 4$\times$300s & 1.1--1.2  & 0.8\\ 
10 Apr 2017    & NTT/SofI & Spec & GB & 0.95--1.64 & 1  & 0.29 & 6.96 & 600 & 4$\times$300s & 1.1--1.2  & 0.8 \\  
29 May 2017  & GTC/OSIRIS & Spec & R300R & 0.48--1.00 & 1 & 0.25 & 7.74 & 240  & 2$\times$600s & 1.7--1.8 & 1.0 \\
22 Mar 2019  & Keck-I/OSIRIS & AO & $K_{p}$ & 1.96--2.27 & -- & 0.01 & -- & --  & 40$\times$8.851s & 1.4--1.5 & 0.9 \\   
\hline
 \noalign{\smallskip}
 \multicolumn{10}{c}{USco1556 A}  \\
\hline
\noalign{\smallskip}
22 Mar 2019  & Keck-I/OSIRIS & AO & $K_{p}$ & 1.96--2.27 & -- & 0.01 & -- & --  & 150$\times$1.475s & 1.4 & 1.0 \\ 

 \hline
\noalign{\smallskip}
 \multicolumn{10}{c}{USco1556 B}  \\
\hline
\noalign{\smallskip} 
 7 May 2007 & UKIRT/WFCAM  & Img & $H, K$ & -- & -- & 0.4 & -- & -- & 10s, 5s & 1.5 & 0.8 \\
  15 Jun 2010  & UKIRT/WFCAM  & Img & $Z$ & -- & -- & 0.4 & -- & -- & 20s & 1.6 & 0.9 \\
  27 Jun 2010  & UKIRT/WFCAM  & Img & $Y, J$ & -- & -- & 0.4 & -- & -- & 20s, 10s & 1.5 & 0.7 \\
  2 Apr 2011 & UKIRT/WFCAM  & Img & $K$ & -- & -- & 0.4 & -- & -- & 10s & 1.5 & 2.0 \\
  1 May 2014 & VISTA/VIRCAM  & Img & $J, Ks$ & -- & -- & 0.34 & -- & -- & 15s, 7.5s & 1.1 &0.9 \\ 
 16 Jun 2018  & VLT/X-shooter & Spec & VIS & 0.55--1.02 & 1.2 & 0.158 & 0.16 & 6500 & 5$\times$280s & 1.0--1.1 & 1.0 \\
 16 Jun 2018  & VLT/X-shooter & Spec & NIR & 1.02--2.48 & 1.2 & 0.248 & 0.77 & 4300 & 5$\times$300s & 1.0--1.1 & 1.0 \\  
 22 Mar 2019  & Keck-I/OSIRIS & AO & $K_{p}$ & 1.96--2.27 & -- & 0.01 & -- & --  & 30$\times$8.851s & 1.4 & 1.0 \\ 
\hline \hline                 
\end{tabular}
\end{table*}

\subsection{Spectroscopy\\ }

We performed low-resolution optical spectroscopy to determine
spectral types and to search for signatures of youth to confirm membership
to the association. 

\subsubsection{GTC/OSIRIS optical spectroscopy of USco1621 B}

USco1621\,B was observed with the Long Slit Spectroscopy mode of the Optical System for Imaging and low-Intermediate-Resolution Integrated Spectroscopy \citep[OSIRIS;][]{OSIRISpaper} spectrograph, mounted on the Gran Telescopio Canarias (GTC) at the Observatorio del Roque de Los Muchachos (ORM) in La Palma, Canary Islands, on 29 May 2017. These observations were performed using the 1.0\arcsec width long slit in parallactic angle, and the R300R grating (resolving power $\sim$240, wavelength range of 4800--10\,000\,\AA). The standard configuration includes a 2$\times$2 binning of the OSIRIS detector. Two individual exposures of 600s were taken at two A--B different nodding positions along the slit.
The average seeing value was around 1.0\arcsec and the airmass of the observations was around 1.7--1.8.

\begin{figure}
   \centering
   \includegraphics[width=9.5cm]{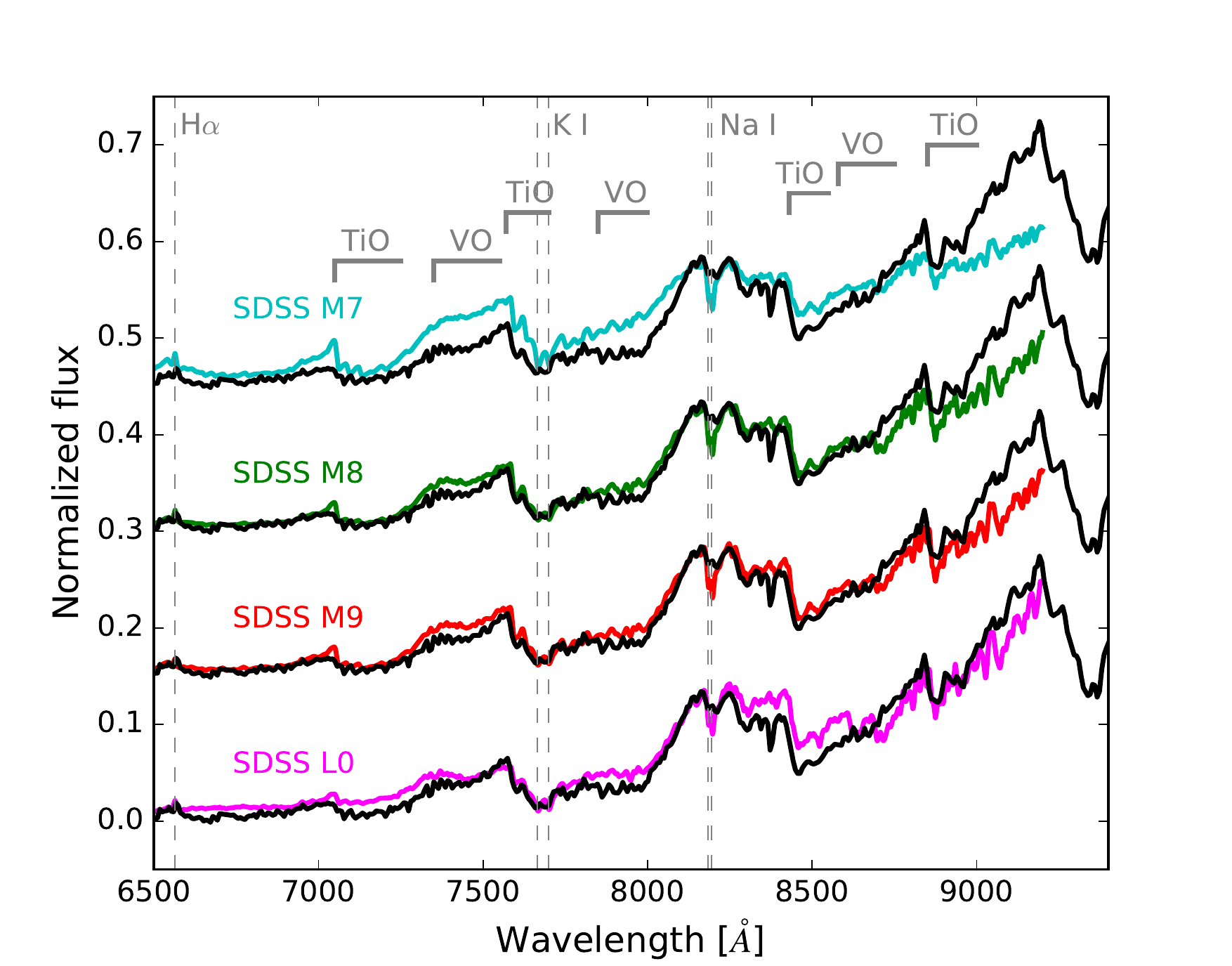}
   \caption{GTC/OSIRIS optical spectrum of USco1621\,B (black) overplotted with SDSS templates (coloured) from \citet{Bochanski2007}, degraded to match the OSIRIS resolution, for comparison. The spectra have been normalised at $\sim$8150\,$\AA$, and shifted by a constant for clarity. The most prominent spectral features are labelled.}
              \label{fig:OPTspectra_SDSS}%
    \end{figure}

\begin{figure}
   \centering
   \includegraphics[width=9.5cm]{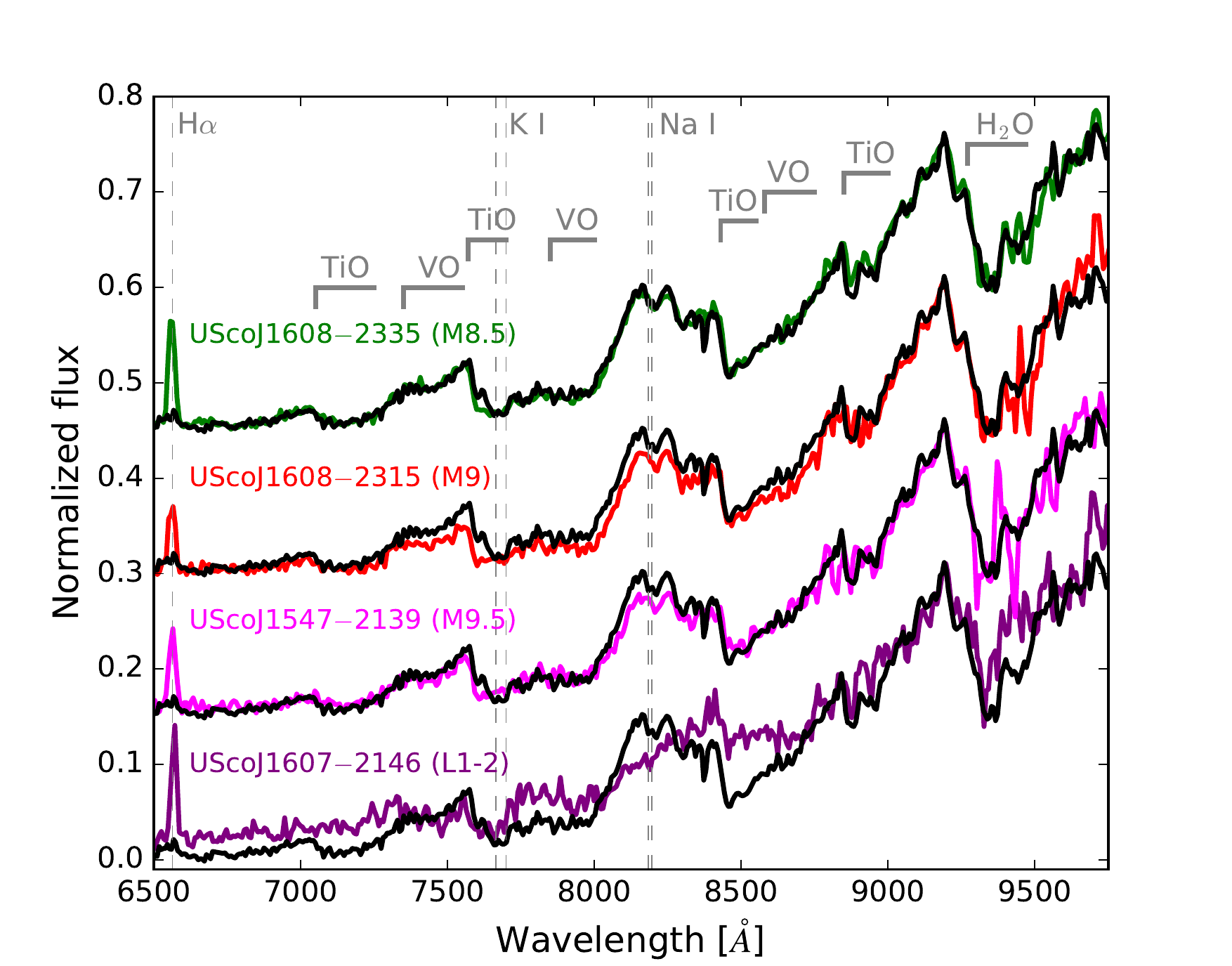}
   \caption{GTC/OSIRIS optical spectrum of USco1621\,B (black) compared to known USco members (coloured) obtained with GTC/OSIRIS using the same instrumental configuration by \citet{Lodieu2018}. The spectra have been normalised at $\sim$9200\,$\AA$, and shifted by a constant for clarity. The most prominent spectral features are labelled.} 
              \label{fig:OPTspectra_USco}%
    \end{figure}

The data reduction was performed using standard routines within the Image Reduction and Analysis Facility (IRAF) environment \citep{IRAFpaper1, IRAFpaper2}. The raw spectral frames were bias subtracted, divided by flat, combined, and extracted using the {\tt{APALL}} routine. The instrumental response correction was performed using a spectrum of the white dwarf spectroscopic standard GD153. This standard was also observed using the broad \textit{z}-band filter to correct the OSIRIS R300R grism second order contamination from the emission at wavelengths 4800--4900 \AA, which appear at wavelengths between 9600--9800 \AA. This contribution is not significant in the case of our secondaries, as these objects barely emit in the bluer wavelengths, but it is noticeable in the white dwarf standard spectrum. The wavelength calibration was performed using HgAr, Ne and Xe arc lamps images. The resulting spectrum is shown in Figures \ref{fig:OPTspectra_SDSS} and \ref{fig:OPTspectra_USco}.

\begin{table}
\caption{Spectral indices for spectral type determination.}             
\label{table:indices}   
 \begin{tabular}{lclc}
 \hline
\noalign{\smallskip}                    
Index & Value & SpT & Index Ref.\tablefootmark{a}\\ 
\hline\hline       
  \noalign{\smallskip}
   &  & USco1621 B &  \\ 
\hline\hline       
  \noalign{\smallskip}
                     
Optical &  &  &  \\ 
\hline
 \noalign{\smallskip} 
 TiO-7140 & 2.66  & M6-M8 & W05 \\
  \noalign{\smallskip} 
 TiO-8465 & 1.89  & M7.5 & S06 \\
 \noalign{\smallskip} 
 TiO5 \tablefootmark{b} & 0.35  & M8.25$\pm$0.50 & K99, CR02 \\
\noalign{\smallskip} 
 VO-a \tablefootmark{b} & 2.51  & L0.0$\pm$0.75/M9.0$\pm$0.5 & R95, CR02  \\
 \noalign{\smallskip} 
 PC3 &  1.97 & M8.5$\pm$0.25 & M99 \\
 \hline
\noalign{\smallskip}                     
Infrared \tablefootmark{c} &  &  &  \\ 
\hline
 \noalign{\smallskip} 
 H$_{2}$O & 1.13  & M9.25$\pm$0.5 & A07 \\
  \noalign{\smallskip} 
 H$_{2}$OD & 0.97  & L0.75 $\pm$0.75 & ML03 \\
  \noalign{\smallskip} 
 H$_{2}$O-1 & 0.69  & L0.25$\pm$1  & S04 \\
  \noalign{\smallskip} 
 H$_{2}$O-2 & 0.89  & M9.25$\pm$0.5  & S04 \\  
 \noalign{\smallskip} 
 FeH &  0.87 & M9.0$\pm$0.5  & S04 \\
 \noalign{\smallskip}
 
 \hline\hline       
  \noalign{\smallskip}
   &  & USco1556 B &  \\ 
\hline\hline       
  \noalign{\smallskip}                    
Optical &  &  &  \\ 
\hline
 \noalign{\smallskip} 
 TiO-7140 & 1.72  & M9--L0 & W05 \\
  \noalign{\smallskip} 
 TiO-8465 & 2.51  & M8--M9 & S06 \\
 \noalign{\smallskip} 
 TiO5 \tablefootmark{b} & 0.52  & M9.0$\pm$0.5 & K99, CR02 \\
\noalign{\smallskip} 
 VO-a \tablefootmark{b} &  2.58 & L1.0$\pm$0.75/M8.5$\pm$0.5 & R95, CR02  \\
 \noalign{\smallskip} 
 PC3 & 2.27  & M9.5$\pm$0.25 & M99 \\
 \hline
\noalign{\smallskip}                     
Infrared \tablefootmark{c} &  &  &  \\ 
\hline
 \noalign{\smallskip} 
 H$_{2}$O & 1.18  & L0.5$\pm$0.5 & A07 \\
  \noalign{\smallskip} 
 H$_{2}$OD & 1.05  & M8.25$\pm$0.75 & ML03 \\
  \noalign{\smallskip} 
 H$_{2}$O-1 & 0.66  & L1.25$\pm$1 & S04 \\
  \noalign{\smallskip} 
 H$_{2}$O-2 & 0.87  &  M9.75$\pm$0.5 & S04 \\  
 \noalign{\smallskip} 
 FeH & 0.83  & L0.0$\pm$0.5  & S04 \\
 \noalign{\smallskip}

\hline  
 
 \hline              
   
\hline                  
\end{tabular}

\tablefoot{
\tablefoottext{a}{References: A07 -- \citet{Allers2007}, CR02 -- \citet{Cruz2002}, K99 -- \citet{Kirkpatrick1999}, M99 -- \citet{Martin1999}, ML03 -- \citet{McLean2003}, R95 -- \citet{Reid1995}, S04 -- \citet{Slesnick2004}, S06 -- \citet{Slesnick2006}, W05 --\citet{Wilking2005}}
\tablefoottext{b}{Calculated using the relations in \citet{Cruz2002}}
\tablefoottext{c}{Results obtained using the \citet{Allers2013} polynomial fits.}
} 
\end{table}

\nocite{Wilking2005; Slesnick2006; Kirkpatrick1999; Reid1995; Martin1999}

\subsubsection{NTT/SofI near-infrared spectroscopy of USco1621 B}

We also performed near-infrared spectroscopy of USco1621\,B to determine its spectral type. The object was observed on 10 April 2017 using the Son Of Isaac (SofI) spectrograph \citep{1998Msngr..91....9M} mounted on the 3.6m New Technology Telescope (NTT) in La Silla Observatory, Chile. The meteorological conditions over the run were good, the airmass value during the exposure was 1.1--1.2 and the average seeing was around 0.8\arcsec (FWHM) in the $J$ filter. We used the low-resolution red and blue grisms, which provide wavelength ranges of 1.53--2.52 $\mu$m and 0.95--1.64 $\mu$m, respectively, and a resolving power of 600, with a nominal dispersion of 10.22 and 6.96\,$\AA$\,pix$^{-1}$, respectively, for a 1.0\arcsec wide slit. Four individual exposures of 300s were taken using an ABBA nodding pattern along the slit, for both red and blue configurations. The B6 telluric standard HIP80126 was observed right after the scientific target for telluric correction at similar airmass. We also acquired dome spectral flat field images and Xe arc-lamp images to perform the flat field correction and the wavelength calibration. 

\begin{figure}
   \centering
   \includegraphics[width=9.5cm]{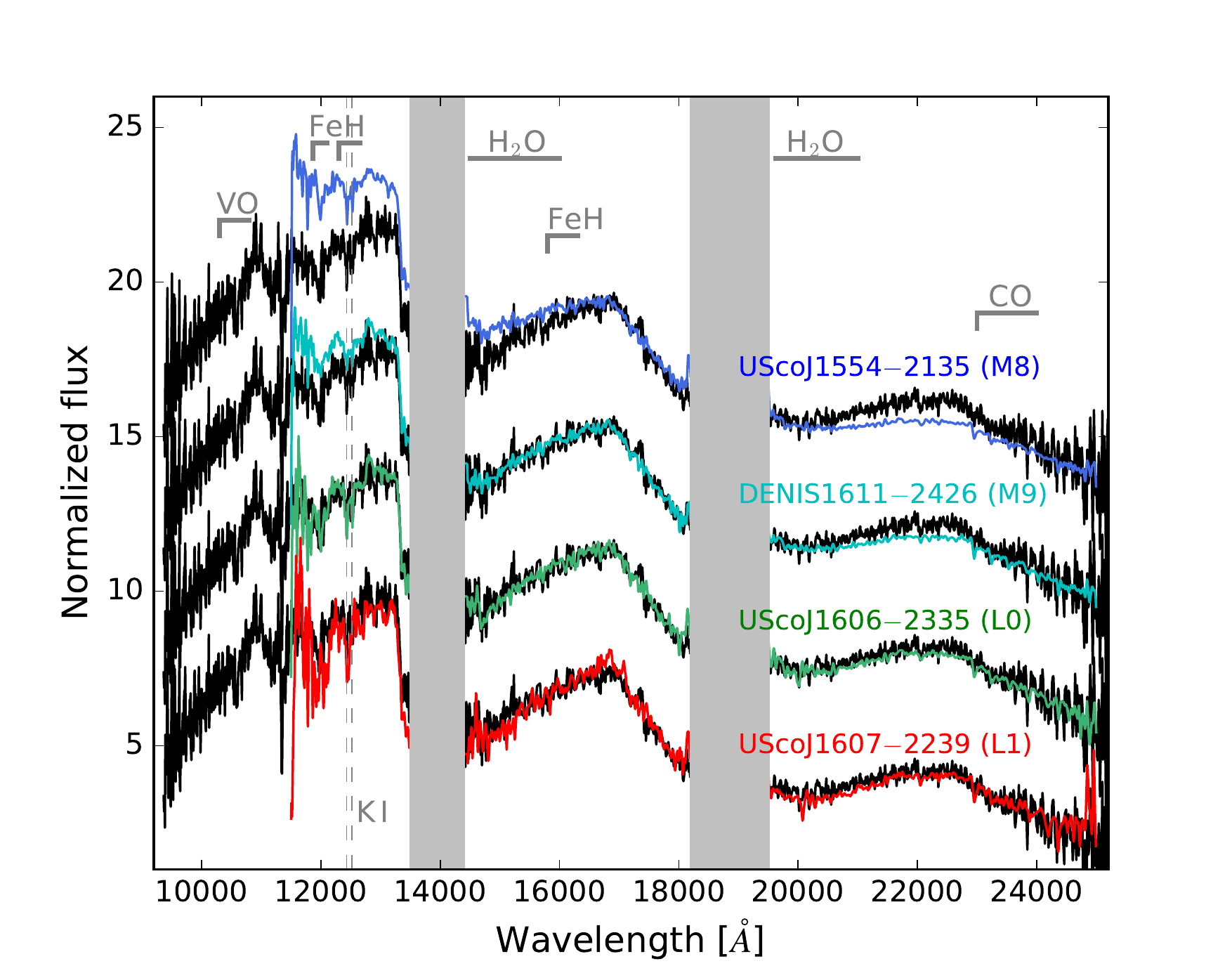}
   \caption{NTT/SofI near-infrared spectrum of USco1621\,B (black) overplotted with USco known members (coloured) obtained with Gemini North/GNIRS by \citet{Lodieu2008}, for comparison. GNIRS spectra were convolved with a gaussian function to match the SofI spectral resolution. The spectra have been normalised in the $H$-band region (wavelength range $\sim$1.5--1.8 $\mu$m), and shifted by a constant for clarity. The regions affected by strong atmospheric telluric absorption bands are marked as grey bands. The most prominent spectral features are labelled.}
              \label{fig:IRspectra}%
    \end{figure}

\begin{figure}
   \centering
   \resizebox{\hsize}{!}{\includegraphics{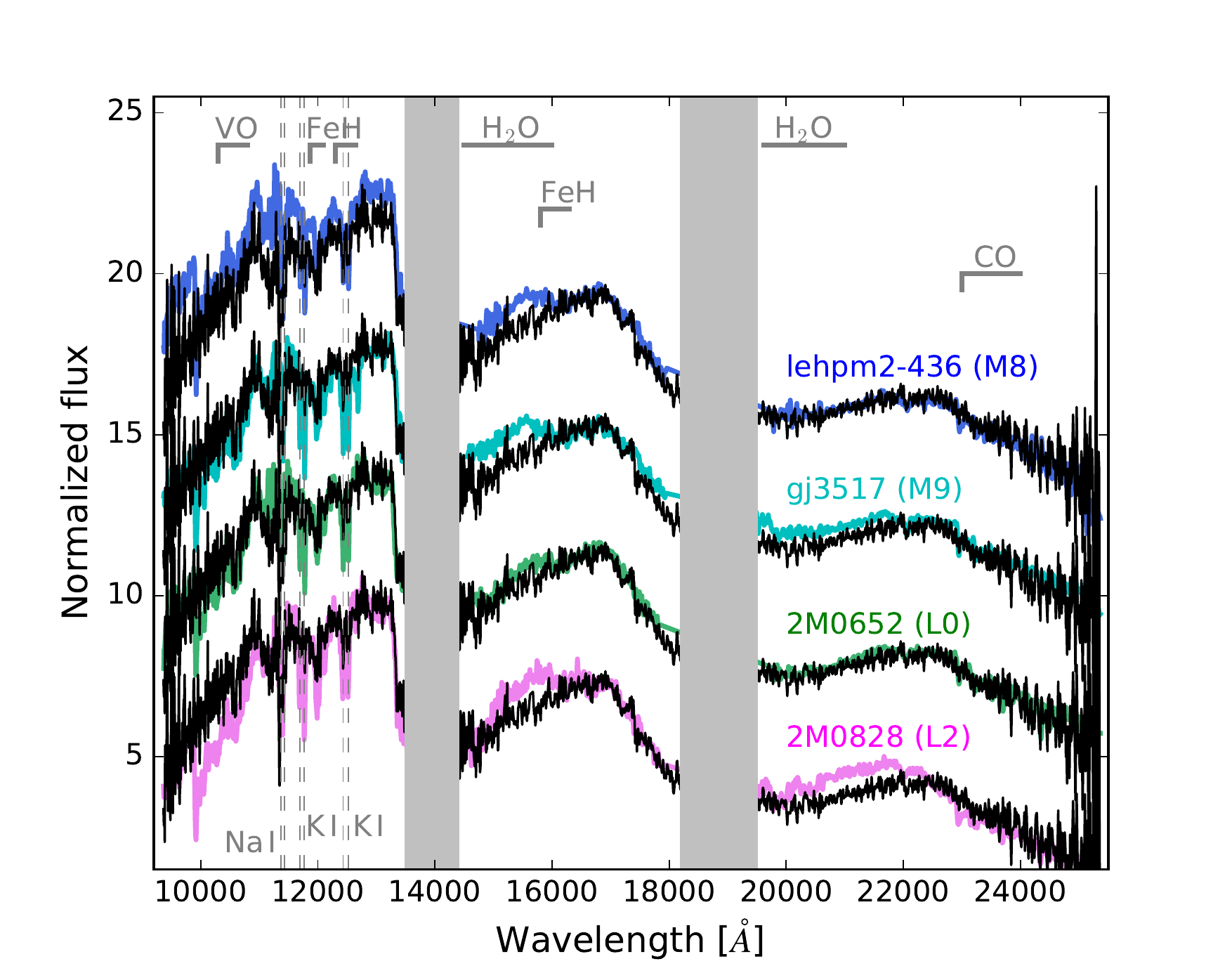}}

   \caption{Same as Fig. \ref{fig:IRspectra}, but comparing USco1621\,B (black) to field dwarf spectra (coloured) observed with NTT/SofI using the same instrumental configuration as the target. }
              \label{fig:IRfieldspectra}%
    \end{figure}

The raw spectroscopic images were flat field corrected, sky subtracted using the A-B and B-A positions to remove the sky emission lines, aligned and combined. Then the spectra were extracted using the IRAF {\tt{APALL}} routine and wavelength calibrated using the Xe arcs. A similar procedure was used for the telluric standard. We manually removed the intrinsic lines of the B6 telluric star spectra. Then, to correct for the instrumental response, the target spectra were divided by the telluric spectra and multiplied by a black body of 14\,500~K. The red and blue spectra were scaled and combined using the overlapping region at $\sim$1.5--1.6 $\mu$m. The obtained spectrum is shown in Figures \ref{fig:IRspectra} and \ref{fig:IRfieldspectra}.

\subsubsection{VLT/X-shooter optical and near-infrared spectroscopy of USco1556 B}

USco1556\,B was observed with the Echelle spectrograph X-shooter \citep{XSHOOTERpaper} mounted on the Very Large Telescope (VLT) at Paranal Observatory (Chile) on 16 June 2018 in service mode.
It was observed simultaneously in the three spectroscopic arms (UVB, VIS and NIR, corresponding to ultra-violet, visible, and near-infrared bands; covering 300--560, 560--1024, and 1024--2480 nm regions, respectively). The target is too faint in the UVB band to obtain a spectrum of adequate signal-to-noise-ratio in a reasonable integration time, so we restrict our analysis to the VIS and NIR bands only. 

The observations were performed using 1.2\arcsec wide slits in both VIS and NIR arms, using an ABBA nodding pattern along the slits. The measured resolving power is $\sim$6500 and $\sim$4300 for VIS and NIR arms, respectively. Five individual exposures of 300s for the near-infrared and 280s for the optical were used. As the readout time is shorter for the NIR arm, it is possible to extend the exposure time in this arm, matching the total observing time in both arms to optimise the observations. The object was near culmination and the average seeing was around 1.0\arcsec. The B8V spectrophotometric standard HD148594 was also observed right after the target at a similar airmass to correct for atmospheric telluric absorption.

\begin{figure}
   \centering
   \includegraphics[width=9.5cm]{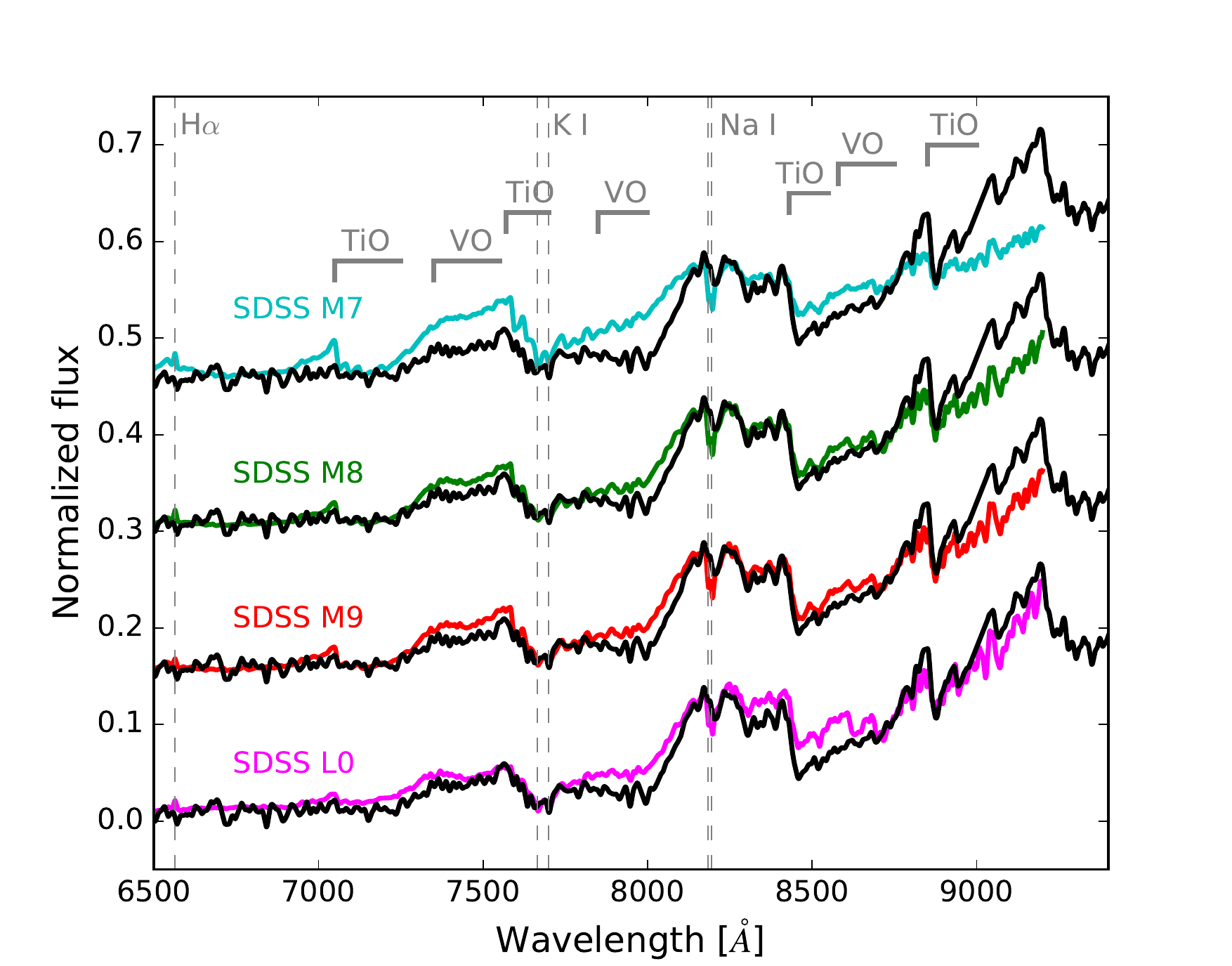}
   \caption{VLT/X-shooter spectrum of USco1556\,B in the VIS band (black) overplotted with SDSS templates (coloured) from \citet{Bochanski2007}, for comparison. For a better visualisation, the X-shooter spectrum has been degraded to the resolution of the SDSS spectra using a Gaussian kernel.
  The spectra have been normalised at $\sim$8150\,$\AA$, and shifted by a constant for clarity. The most prominent spectral features are labelled.} 
              \label{fig:MortyOPTspectra_SDSS}%
    \end{figure}

\begin{figure}
   \centering
   \includegraphics[width=9.5cm]{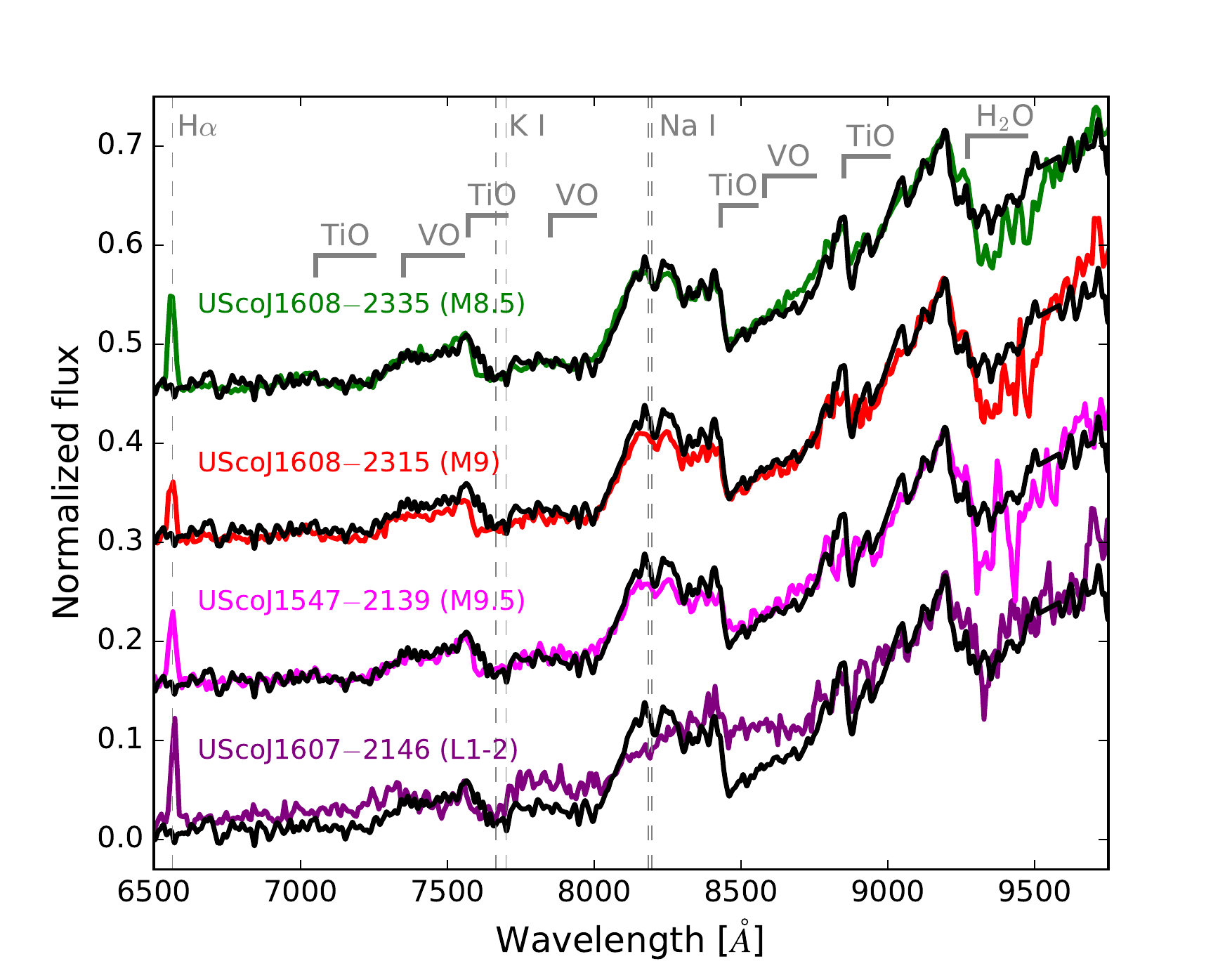}
   \caption{VLT/X-shooter spectrum of USco1556\,B in the VIS band (black) overplotted with USco known members (coloured) obtained with GTC/OSIRIS by \citet{Lodieu2018}. X-shooter spectrum has been convolved using a gaussian function to the OSIRIS resolution. The spectra have been normalised at $\sim$9200\,$\AA$, and shifted by a constant for clarity. The most prominent spectral features are labelled.}
              \label{fig:MortyOPTspectra_USco}%
    \end{figure}

The data reduction was performed using the ESO REFLEX environment \citep{ESOREFLEXpaper}, which is based on the open source workflow engine Kepler, and IRAF. We used ESO REFLEX to obtain the combined, wavelength calibrated, and flux-calibrated 2D spectra. The X-shooter pipeline \citep{XSHOOTERpipeline} performs bias and dark subtraction; flat-fielding, non-linearity and instrument flexure correction; instrument response and wavelength calibration; and spectral orders combination. Then the spectra were extracted using IRAF {\tt{APALL}} routine. To perform the telluric correction in the near-infrared, we manually removed the characteristic spectral lines of the telluric star spectrum, divided the spectrum of USco1556\,B by the telluric standard spectrum, and then multiplied the result by a black body of 12\,500~K. For the optical, we also performed a telluric correction of the spectrum. In this case we also used the IRAF {\tt{CONTINUUM}} routine to flatten the telluric spectrum before using it to divide the target optical spectrum. The optical and infra-red spectra of USco1556\,B are shown in Figures \ref{fig:MortyOPTspectra_SDSS} to \ref{fig:IRfieldspectraMorty}.

\subsubsection{Spectral classification in the optical}

For the spectral classification in the optical, we have compared the spectrum of USco1621\,B and USco1556\,B with spectral templates of the Sloan Digital Sky Survey \citep[SDSS; ][]{SDSSpaper} obtained by \citet{Bochanski2007}. Figures \ref{fig:OPTspectra_SDSS} and \ref{fig:MortyOPTspectra_SDSS} show the GTC/OSIRIS optical spectrum of USco1621\,B and the VLT/X-shooter optical spectrum of USco1556\,B together with SDSS standards from M7 to L0. For the comparison, we degraded the resolution of the SDSS and X-shooter spectra by convolving them with a Gaussian kernel, to match the resolution of the OSIRIS spectra. From these comparisons we can see that the depth of the TiO and VO molecular bands match better the M8--M9 spectral types in both cases, but the pseudo-continua follow the spectral energy distribution of the L0 standard. This is probably due to the difference in age of the objects we are trying to compare. The SDSS standards are built from field-age objects, while our targets are much younger, and oxides are more prominent in low gravity atmospheres \citep{Martin1996, ZapateroOsorio1997, Allers2013, ZapateroOsorio2017}. This difference in age can also be appreciated in other youth features such as very weak Na~I and K~I doublets absorption lines (see section 3.1.6 for more details). 

We have also compared the optical spectra of USco1621\,B and USco1556\,B with spectra of known low-mass USco members. Figures \ref{fig:OPTspectra_USco} and \ref{fig:MortyOPTspectra_USco} show the optical spectrum of these targets together with 4 USco objects with spectral types M8.5, M9, M9.5 and L1--L2 obtained with GTC/OSIRIS \citep{Lodieu2018}, for comparison. The best fits are obtained with the M8.5 object UScoJ16083049$-$2335110 for both secondaries.

In addition, we computed a set of typical spectral indices used for spectral classification, whose results are shown in Table~\ref{table:indices}. The VO-a index \citep{Reid1995, Cruz2002} is right in the frontier of the two polynomial fits provided by \citet{Cruz2002}, so we include the results of the two fits.
We caution that oxide bands are gravity-sensitive features, and therefore, these indices may not be suitable for spectral type determination at young ages. As a result of all the visual comparisons and spectral-index determinations, we finally adopted a  spectral type of M8.5$\pm$0.5 in the optical for both USco1621\,B and USco1556\,B. We estimated the error bars taking into account that our spectra deviate noticeably from the reference spectra having spectral types 1 subclass earlier or later.

\subsubsection{Spectral classification in the near-infrared}

  \begin{figure}
   \centering
   \includegraphics[width=9.5cm]{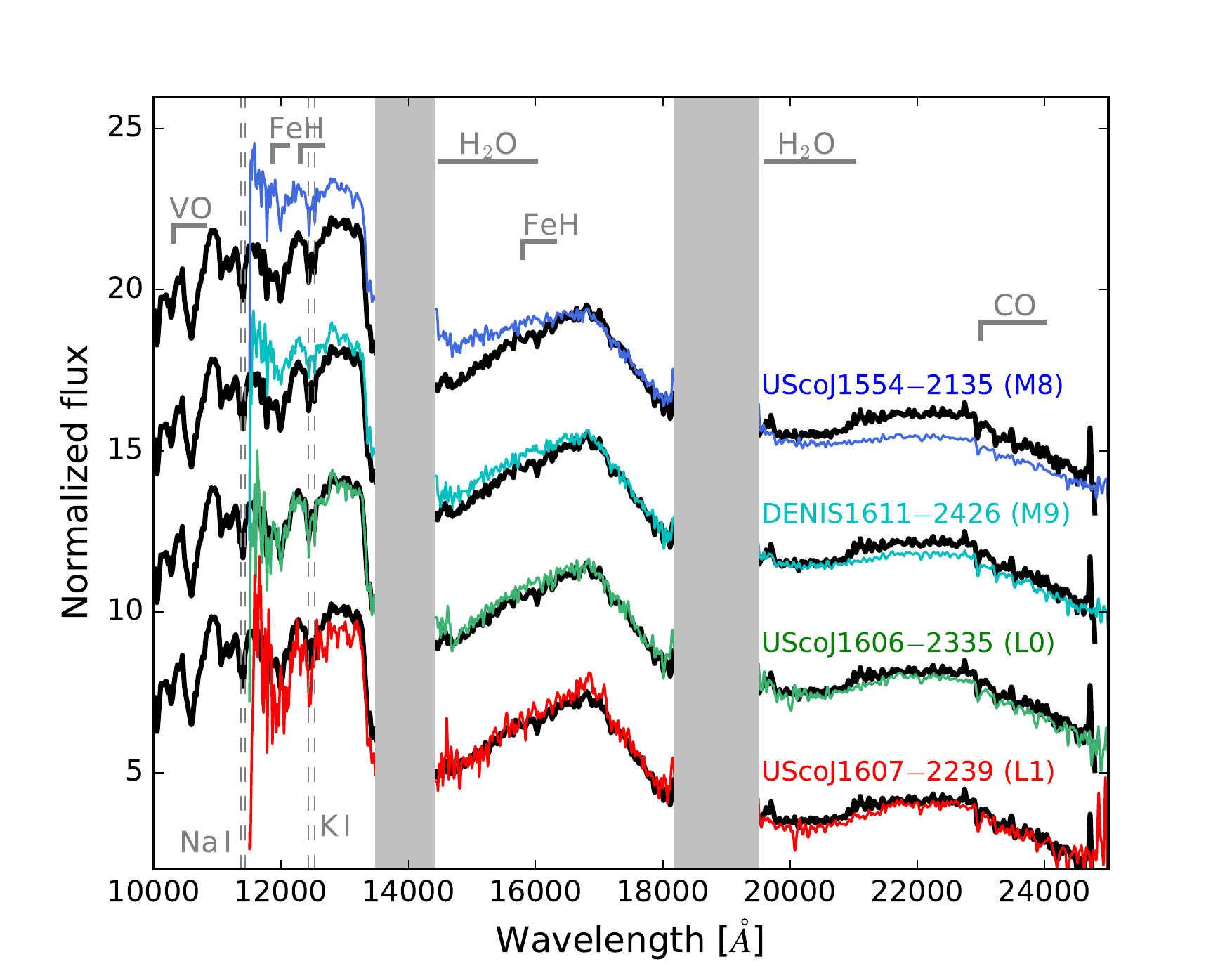}
   \caption{VLT/X-shooter spectrum of USco1556\,B in the NIR band (black) overplotted with USco known members (coloured) obtained with Gemini North GNIRS by \citet{Lodieu2008}, for comparison. The X-shooter spectrum has been smoothed using a Gaussian kernel. GNIRS spectra were also degraded by gaussian convolution to match the same spectral resolution. The spectra have been normalised in the $H$-band region (wavelength range $\sim$1.5--1.8 $\mu$m), and shifted by a constant for clarity. Zones affected by atmospheric telluric bands are marked as grey bands. The most prominent spectral features are labelled.}
              \label{fig:MortyIRspectra}%
    \end{figure}

 \begin{figure}
   \centering
   \includegraphics[width=9.5cm]{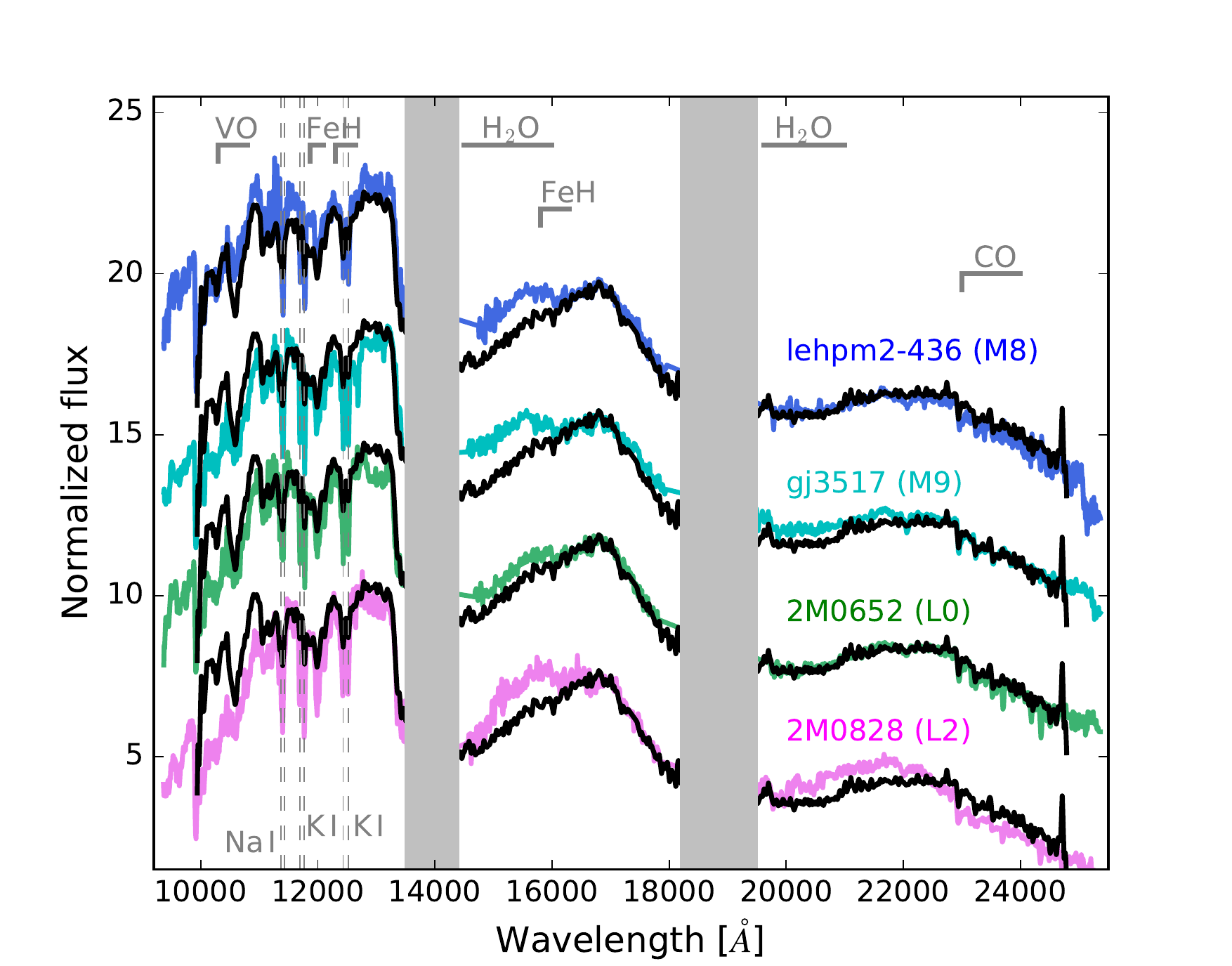}
   \caption{Same as in Fig. \ref{fig:MortyIRspectra}, but here comparing USco1556\,B (black) to field stars spectra (coloured) obtained with NTT/SofI. }
              \label{fig:IRfieldspectraMorty}%
    \end{figure}

To perform the spectral classification in the near-infrared region, we compared the  spectra of USco1621\,B and USco1556\,B with spectra from previously-known USco members observed with the Gemini Near-Infrared Spectrograph GNIRS \citep{GNIRSpaper} on Gemini North by \citet{Lodieu2008}. Figures \ref{fig:IRspectra} and \ref{fig:MortyIRspectra} show the near-infrared spectrum of USco1621\,B and USco1556\,B together with spectra of M8, M9, L0 and L1 USco members: UScoJ155419$-$213543, DENIS161103-242642, UScoJ160606$-$233513, and UScoJ160727$-$223904, respectively. We degraded the GNIRS and X-shooter spectra to match the resolution of SofI. The spectral energy distributions of USco1621\,B and USco1556\,B are both very similar to UScoJ160606$-$233513, which is classified as an L0 in the infrared, although they are slightly redder in the K band. This difference is more noticeable for USco1556\,B, and also the peak in the \textit{H} band is more pronounced for this secondary, which may indicate a slightly later spectral type. When comparing with the L1 spectral type UScoJ160727$-$223904 we find a larger discrepancy in the spectral energy distribution, mostly in the \textit{J} band.

We also compared the near-infrared spectrum of both secondaries with several field-age objects with spectral types M8--L2, observed with NTT/SofI using the same instrumental configuration. These spectra were obtained and reduced following a similar procedure to USco1621\,B. The comparisons are shown in Fig. \ref{fig:IRfieldspectra} and Fig. \ref {fig:IRfieldspectraMorty}. The spectral energy distributions are very similar to the object 2MASS 0652, which is classified as an L0. This typing coincides with that obtained from the young references. We can also distinguish some youth features in our targets, such as weak K~I alkali lines and the triangular shape in the \textit{H} band, which are not present in the spectra of the field counterparts \citep[e.g.,][]{Lucas2001, Gorlova2003, McGovern2004, Allers2013}.

We also computed several useful spectral indices in the near-infrared for the spectral classification. We used the water indices H$_{2}$O \citep{Allers2007}, H$_{2}$OD \citep{McLean2003}, H$_{2}$O-1, and H$_{2}$O-2 \citep{Slesnick2004}, and the FeH index from \citet{Slesnick2004}. The results obtained are shown in Table \ref{table:indices}. We used the polynomial fits described in \citet{Allers2013}, which relate the water indices to the corresponding spectral type in the optical for typing our targets. Hydrides like FeH are weak in low-gravity atmospheres. Therefore, this index may not be suitable for spectral type determination as it is atmospheric gravity dependent.
Using these comparisons and indices, we finally assigned a spectral type of L0$\pm$0.5 to USco1621\,B and of L0.5$\pm$0.5 to USco1556\,B in the near-infrared. We estimated the error bars taking into account that our spectra do not resemble the reference spectra with spectral types 1 subclass earlier or later than USco1621\,B and USco1556\,B.

The spectral type determination in the near-infrared does not perfectly coincide with that estimated in the optical. This is not an unusual behaviour for young low-mass objects, and similar discrepancies have been reported earlier in USco and other young clusters and associations \citep[e.g.][]{Luhman1999, Barrado2001, Luhman2003, Stauffer2003, Pecaut2016, Lodieu2018}.

\begin{table}
\begin{center}
\caption{Gravity-dependent indices.}             
\label{table:lgindices}   
 \begin{tabular}{p{0.14\linewidth}lccl}
 \hline
\noalign{\smallskip}                    
Index & Value & Gravity\tablefootmark{a} & Index Ref.\tablefootmark{b} \\ 
\hline\hline       
  \noalign{\smallskip}
    \multicolumn{4}{c}{USco1621 B}  \\ 
\hline\hline       
 
 \noalign{\smallskip} 
 H-cont & 0.95  & $\gamma$ & AL13 \\
 \noalign{\smallskip}
 CH$_{4}$-H & 1.07  & $\beta$  & B06 \\
 \noalign{\smallskip}
 VO$_{z}$ & 1.11  & \,\,\,\,?\,\tablefootmark{c} & AL13 \\
 \noalign{\smallskip}
 H$_{2}$O-K & 0.95  & $\gamma$ & B06 \\

 \hline\hline       
  \noalign{\smallskip} 
   \multicolumn{4}{c}{USco1556 B}  \\ 
\hline\hline

 \noalign{\smallskip} 
 H-cont & 0.96  & $\gamma$ & AL13 \\
 \noalign{\smallskip}
 CH$_{4}$-H & 1.12  & $\gamma$ & B06 \\   
 \noalign{\smallskip}
 VO$_{z}$ & 1.26  & $\gamma$ & AL13 \\
 \noalign{\smallskip}
 H$_{2}$O-K & 0.94  & $\gamma$ & B06 \\

\hline  
 
 \hline              
   
\hline                  
\end{tabular}
\end{center}
\tablefoot{
\tablefoottext{a}{Gravity indices as described in \citet{Kirkpatrick2005} and \citet{Cruz2009}}. Assigned following the plots in Fig. A1 of Appendix A in \citet{Lodieu2018}.
\tablefoottext{b}{References: AL13 -- \citet{Allers2013}, B06 -- \citet{Burgasser2006}}
\tablefoottext{c}{This value is not conclusive for this NIR spectral type, and could indicate either a $\gamma$, $\beta$ or Field gravity subclass.}
} 

\end{table}

\subsubsection{Low-gravity signatures}

There are several spectral features from optical and near-infrared observations that can be used as estimators of the surface gravity of an ultra-cool dwarf (T$_{\mathrm{eff}}$ typically below 2500~K). In the infrared, we can distinguish weaker alkali lines, particularly 1.169 / 1.177 $\mu$m and $1.243 / 1.252  ~\mu$m K~I doublets, and a triangular shape of the spectrum in the \textit{H} band due to strong water absorption in both sides of the band. Figures \ref{fig:IRfieldspectra} and  \ref{fig:IRfieldspectraMorty} show the comparison of the companions with field objects in the near-infrared where these features can be clearly observed.

In the optical, we can also see weak alkali lines, especially the 8183 / 8195\,$\AA$ Na~I doublet, and also the 7665 / 7699\,$\AA$ K~I doublet. Figures \ref{fig:OPTspectra_SDSS} and \ref{fig:MortyOPTspectra_SDSS} show the comparison between the young companions and several field dwarf objects in the optical. We note that these doublets may be affected by telluric absorption, and no telluric correction is applied to the GTC/OSIRIS optical spectrum of USco1621~B. The doublets in this object may therefore be even weaker than they appear. 

We did not find H-alpha in emission, which is a common indicator of youth, in any of the two companions. The fraction of objects showing H-alpha emission has its maximum occurrence rate at late-M and early-L spectral types, and declines for mid-L and later spectral types \citep[e.g. ][]{Schmidt2015, Pineda2016}. A similar behaviour is also found in young objects of the USco association, as can be seen in \citet{Lodieu2018}. However, the lack of H-alpha emission is not a peculiar feature for these objects, as USco members with similar spectral types may have H-alpha in emission or not \citep[see Figures 6 and 9 in ][]{Lodieu2018}. There is also the possibility that the lack of H-alpha emission in some low-mass USco members spectra may be due to variability of the chromospheric activity \citep{Petrus2019}.

We also computed gravity-sensitive indices to assess the young nature of our targets and, consequently, their likely membership in the USco association. \citet{Lodieu2018} analysed the most suitable indices for gravity characterisation, and concluded that H-cont, CH$_{4}$-H, FeH-H, VO$_{z}$ , and H$_{2}$O-K \citep{Burgasser2006, Allers2013} are the most suitable gravity tracers for low resolution spectra.
Table \ref{table:lgindices} shows the measured indices and the gravity subtype. The gravity subtypes $\beta$ and $\gamma$ denote intermediate and very low surface gravity, respectively \citep{Kirkpatrick2005, Cruz2009}. To estimate these, we followed the positions in the plots presented in Fig. A1 of the Appendix A in \citet{Lodieu2018}. 
Taking into account both the visual comparison and computed indices, we classified both companions as very low-gravity objects compatible with a very young age, like that of USco.

\begin{figure*}
   \centering
  \includegraphics[width=8.7cm]{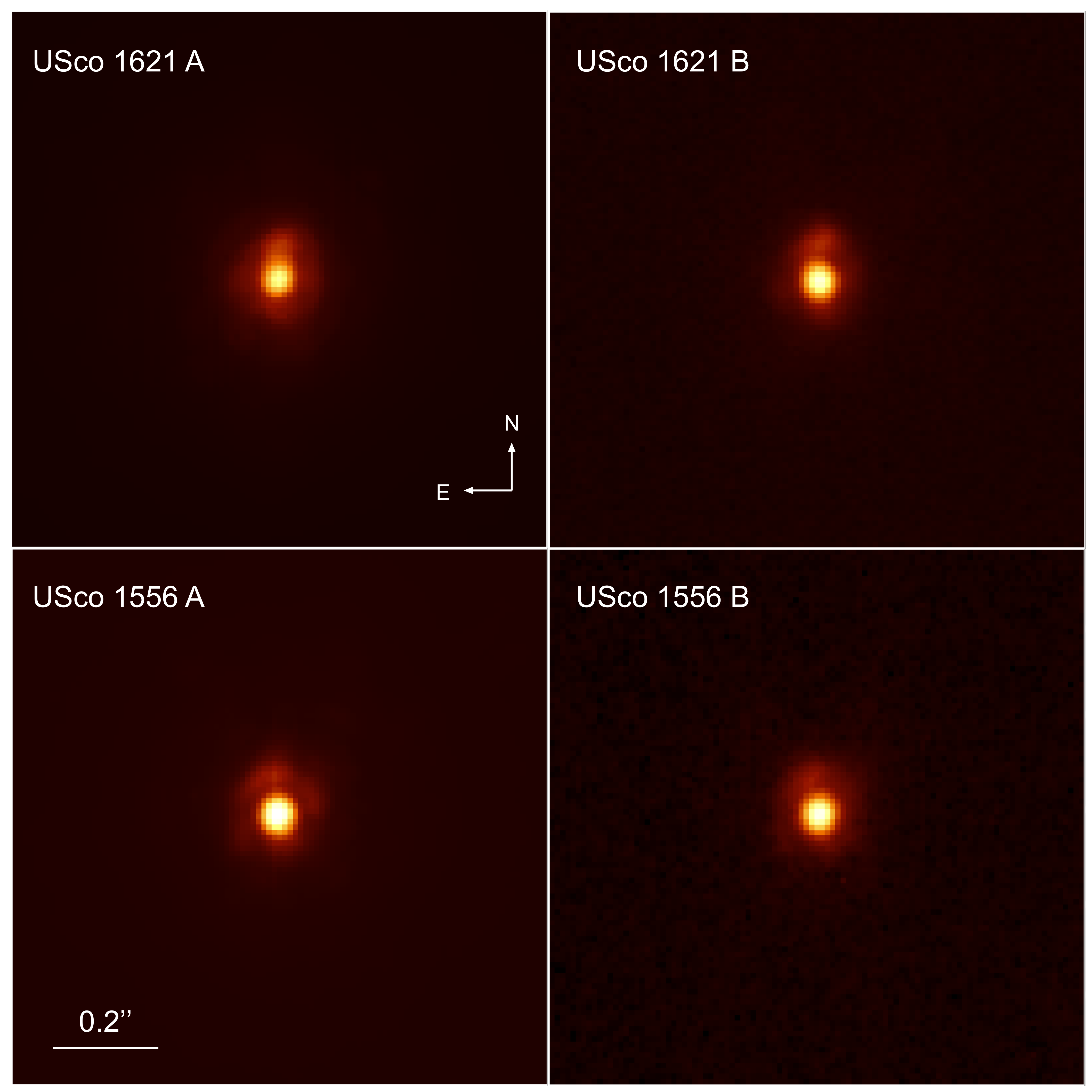}
   \includegraphics[width=9.6cm]{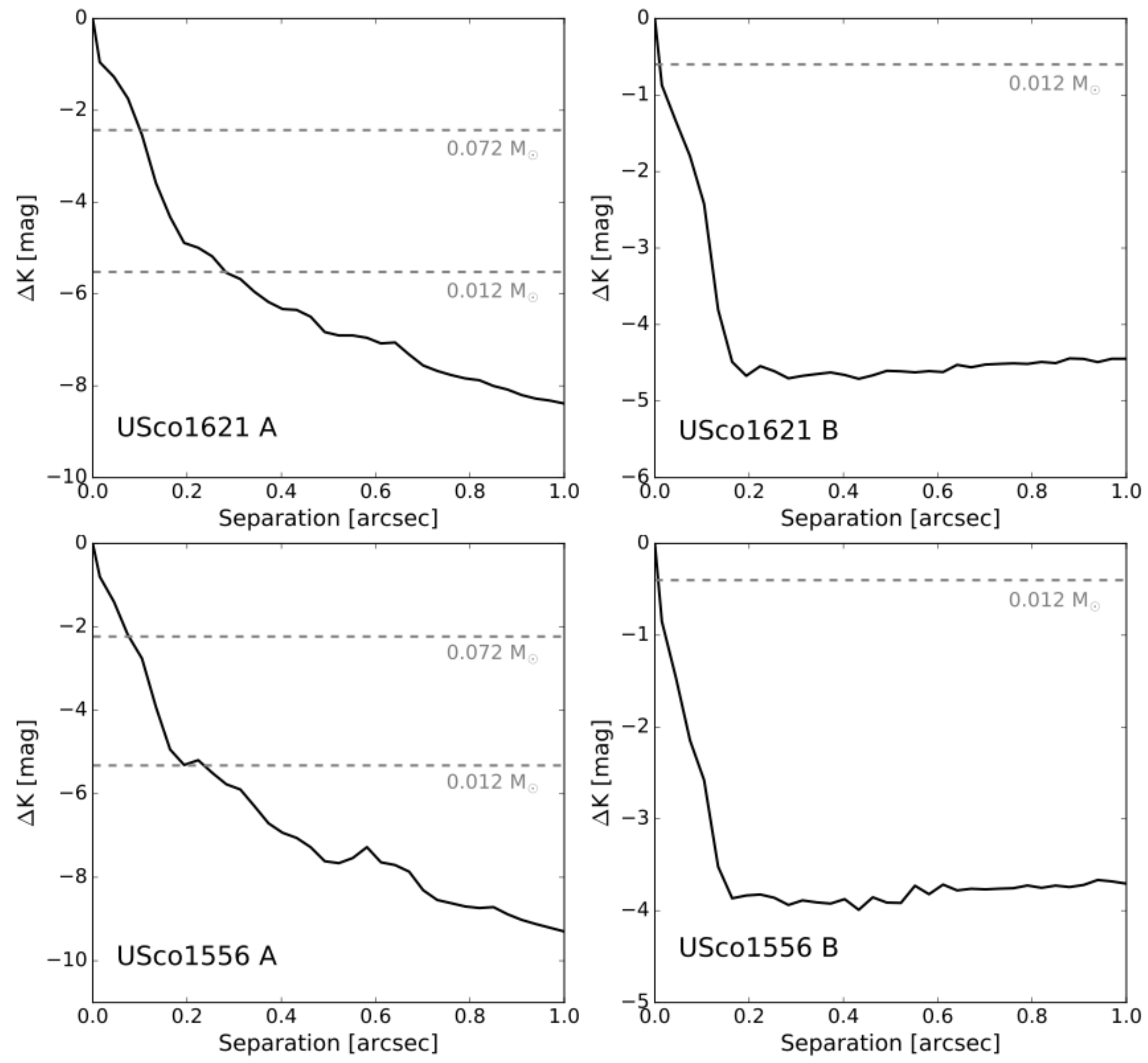}
   \caption{\textit{Left panels:} Keck-I/OSIRIS Adaptive Optics images of the primaries and secondaries of USco1621 and USco1556 systems, in the $K_{p}$ filter. The field-of-view of the images is 1\arcsec $\times$ 1\arcsec and the orientation is North up and East to the left. Contrast scale of the images has been adjusted individually. \textit{Right panels:} Contrast curves for the primaries and secondaries of the USco1621 and USco1556 systems for a 3$\sigma$ detection limit. The expected differential magnitudes for objects at the theoretical mass limits for hydrogen and deuterium burning are marked by dashed lines.}
              \label{fig:KeckAO}%
    \end{figure*}

\subsection{Keck-I/OSIRIS Adaptive Optics imaging}

We observed both the primaries and secondaries of the USco1621~AB and USco1556~AB systems at the Keck I Telescope on Maunakea, Hawai'i, using the OH-Suppressing Infra-Red Imaging Spectrograph (OSIRIS) \citep{OSIRISKeckpaper} and the Keck Adaptive Optics (AO) system \citep{KeckAOpaper,KeckAOpaper2}, on 22 March 2019. We used the Laser Guide Star (LGS) mode with the OSIRIS imaging arm \citep{OSIRISKeckdetectorpaper}. The field of view of the imager is 20$\arcsec\times$20$\arcsec$, and the pixel scale is $\sim$0.010$\arcsec$.

We observed our targets using the $K_{p}$ filter ($\lambda_c$=2.144 $\mu$m, $\Delta\lambda$=0.307 $\mu$m). The airmass was between 1.4--1.6 and the raw seeing was between 0.9$\arcsec$--1.0$\arcsec$. We used a 5-point dither pattern, with individual exposures of 1.475s for the primaries and 8.851s for the secondaries. Raw images were sky-subtracted, aligned, and combined using IRAF. Figure \ref{fig:KeckAO} shows the reduced images of both primaries and secondaries. We also show the contrast curves for the primaries and the secondaries, considering a 3$\sigma$ detection limit. The contrast curve of the primaries is mostly limited by the contribution of the point spread function (PSF) wings of the bright central object, while for the secondaries it is mostly limited by the background noise.

We do not find any sign of binarity for any of the targets within our contrast limits. For the primaries, we can discard the presence of any low-mass stellar companion at angular separations larger than 0.10\arcsec ($\sim$14 AU) and 0.07\arcsec ($\sim$10 AU) from USco1621~A and USco1556~A, respectively; and the presence of brown dwarf companions at separations greater than 0.3\arcsec ($\sim$40 AU). For the secondaries, we can discard the presence of objects with magnitude differences up to 4.4 and 3.7 mag in $K$ (which correspond to masses higher than $\sim$5 M$_{\mathrm{Jup}}$ for the age of USco) at separations greater than 0.2\arcsec ($\sim$28 AU) from USco1621~B and USco1556~B.

\section{Luminosity, effective temperature and mass estimations}

\begin{figure*}
   \centering
   \includegraphics[width=18.5cm]{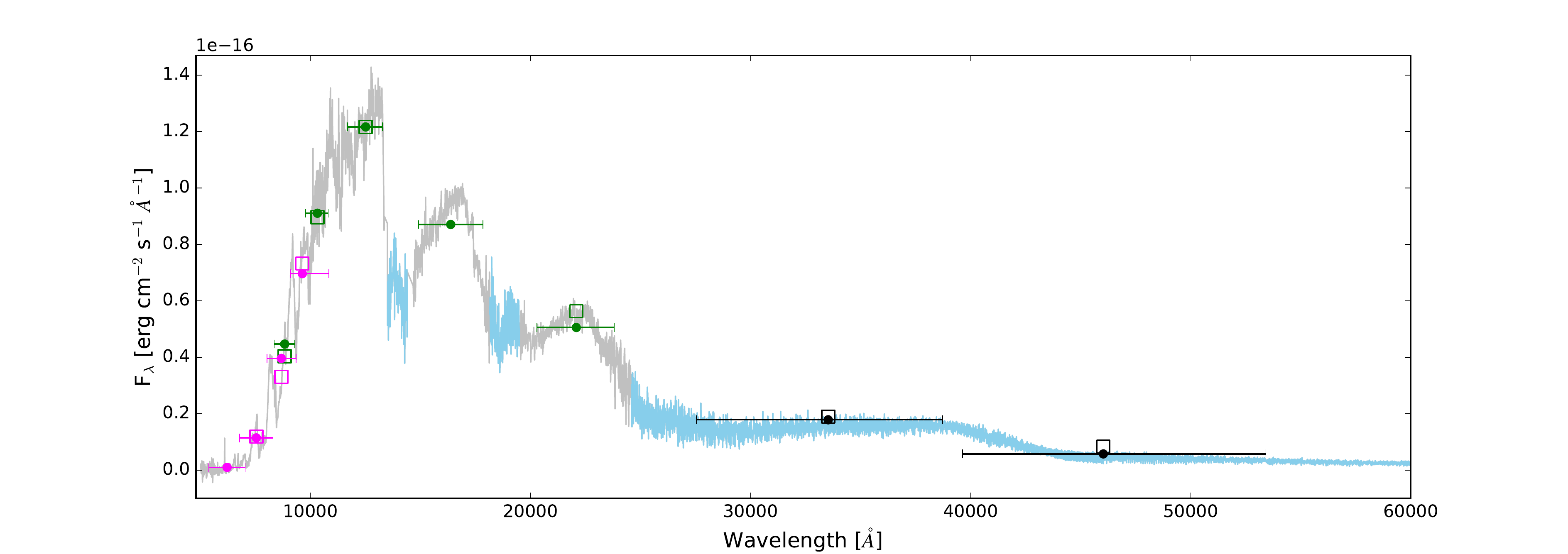}
   \includegraphics[width=18.5cm]{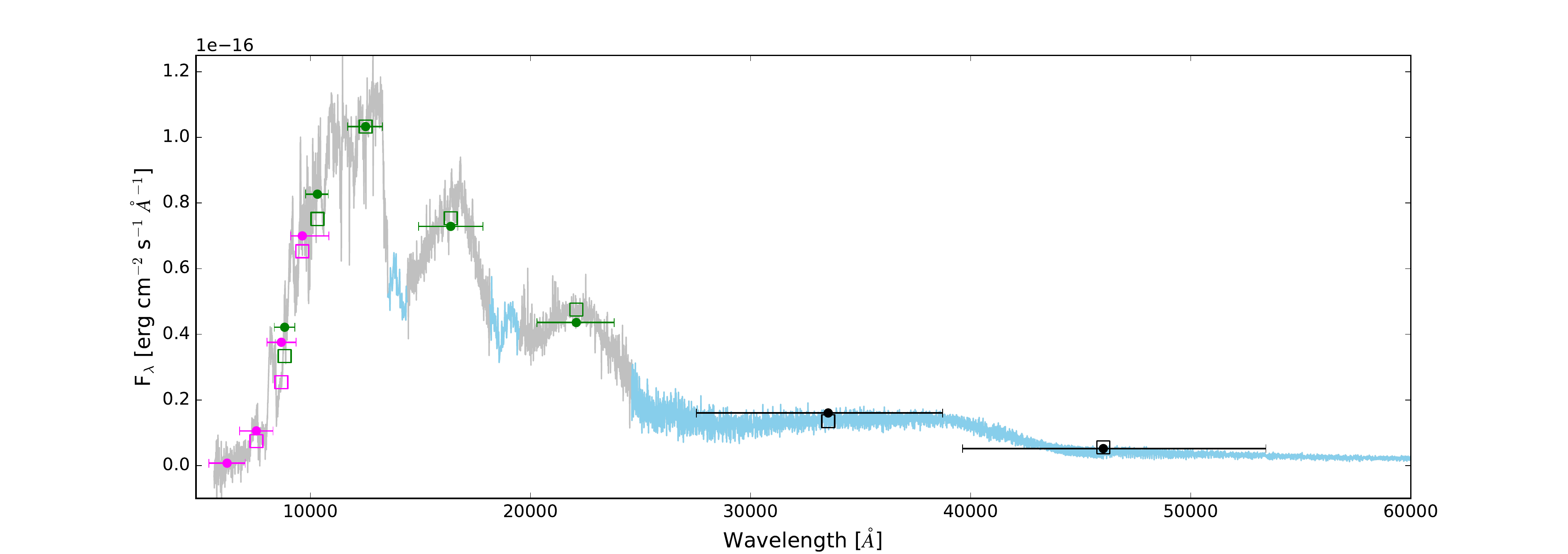}
   \caption{Optical + Infrared spectra of USco1621\,B (upper panel) and USco1556\,B (lower panel), in grey, completed with a $T_{\mathrm{eff}}$=2000~K and log($g$)=4.0 BT-Settl model spectrum from \citet{Allard2012, Baraffe2015}, in light blue. Available photometry from Pan-STARRS (pink open squares), UKIDSS (green open squares), and AllWISE (black open squares) is added. Vertical error bars of the photometry are smaller than the markers. Filled circles correspond to the flux values obtained through direct integration of the spectra in the same filter bands. Horizontal error bars represent the filter bandwidths. }
              \label{fig:SED}%
    \end{figure*}

As a first approach to obtain the luminosity of USco1621\,B and USco1556\,B, we calculated their absolute bolometric magnitudes from the VHS \textit{Ks}-band photometry, bolometric corrections, and assuming the same heliocentric distances as the primaries, measured by $Gaia$ DR2. The bolometric corrections in the \textit{Ks} filter have been derived from the polynomial relations for young objects of the same spectral type from \citet{Filippazzo2015}.

For USco1621\,B, with an optical spectral type of M8.5$\pm$0.5 and a distance of 138$\pm$1pc, we obtained a luminosity of $\log{(L/L_{\odot})}$=$-2.96\pm$0.08. For USco1556\,B, considering an optical spectral type of M8.5$\pm$0.5 and a distance of 141$\pm$2 pc, we obtained a luminosity of $\log{(L/L_{\odot})}=  -3.02\pm$0.09. The errors have been estimated taking into account the photometric error bar from VHS, the parallax error from $Gaia$ DR2, the uncertainty in the spectral type determination and the error in the bolometric corrections.

We also computed the luminosity of the primaries in a similar way. We used their 2MASS photometry in the \textit{J} band and their spectral types from \citet{Rizzuto2015}. The distances were obtained from $Gaia$ DR2, and the bolometric corrections for the \textit{J} band from \citet{PecautMamajek2013}. We obtain, for USco1621~A and USco1556~A, luminosities of $\log{(L/L_{\odot})}$=$-$1.02$\pm$0.03 and $-$1.09$\pm$0.04, respectively.

\begin{figure*}
   \centering
   \includegraphics[width=15cm]{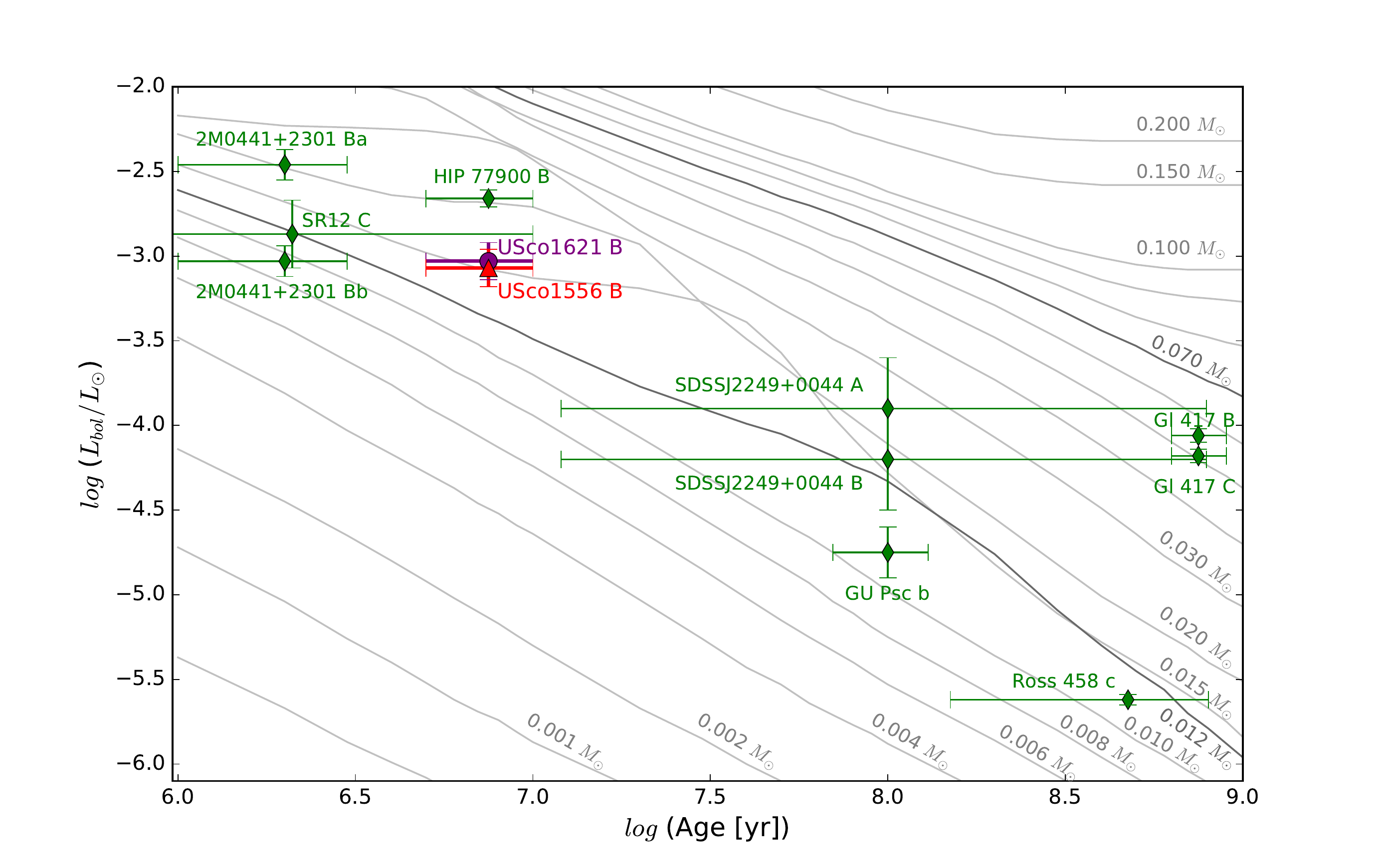}
   \caption{Log${(L/L_{\odot})}$ vs. age diagram for USco1621\,B (purple circle) and USco1556\,B (red triangle), together with Ames-COND model isomasses from \citet{Allard2001} and \citet{Baraffe2003}. Theoretical hydrogen- and deuterium-burning mass limits are marked in dark grey. Other known young wide substellar companions with separations above 1000 AU from \citet{Kirkpatrick2001, Bouy2003, Allers2010, Goldman2010, Todorov2010, Kuzuhara2011, Aller2013, Dupuy2014, Naud2014, Bowler2015} (green diamonds) are also included.}
              \label{fig:Lum_vs_age}%
    \end{figure*}

We also derived the luminosity of the secondaries by integrating the total flux from the optical and infrared spectra. We combined the optical and near-infrared spectra by scaling them, using the mean values of the flux in the overlapping region. Then we flux-calibrated our spectra using the UKIDSS GCS photometry in the \textit{J} band and the integration of the flux of our spectra in this band obtained using the IRAF {\tt{SBANDS}} routine, as our targets have their maximum emission in this wavelength range. We completed the missing mid-infrared wavelengths and replaced regions of the spectra affected by strong telluric absorption using a BT-Settl model spectrum from \citet{Allard2012, Baraffe2015}. We chose the 2000~K and log(g)=4.0 model, as it better reproduces the spectral energy distribution in the near-infrared and the WISE \textit{W1}, \textit{W2} bands. For wavelengths longer than 100\,000~$\AA$, we extrapolated using a Rayleigh-Jeans tail approximation ($f=c\cdot\lambda^{-4}$) although its contribution to the total flux is negligible. For wavelengths shorter than $\sim$5000~$\AA$, the flux is very low (close to zero) and its contribution is also negligible. 

Figure \ref{fig:SED} shows the spectra for USco1621\,B and USco1556\,B together with the available photometry from Pan-STARRS, UKIDSS, and AllWISE. We also show the integrated flux in the same bands, obtained using IRAF {\tt{SBANDS}} routine and convolving with the Pan-STARRS, UKIDSS, and AllWISE filter profiles, which were provided by the Spanish Virtual Observatory Filter Profile Service \citep{SVOFilter}. By integrating the full spectra, we calculated the bolometric luminosity using the $Gaia$ DR2 distances of the primaries. We obtain a bolometric luminosity of $\log{(L/L_{\odot})}$=$-$3.03$\pm$0.11 and $-$3.07$\pm$0.11 for USco1621\,B and USco1556\,B, respectively. These values are in good agreement, at 1$\sigma$ level, with the ones obtained from the photometry alone. The quoted errors include the spectral noise, the uncertainty in the flux calibration, the error given by the selection of slightly different BT-Settl model-spectra for the mid-IR wavelengths, and the error in the distance. 

We estimated the mass and effective temperature of the objects interpolating the BT-Settl theoretical evolutionary models \citep{Allard2012, Baraffe2015} for the corresponding bolometric luminosities, and for an age of the USco association of 5--10 Myr. For USco1621~A and USco1556~A, we obtain masses of 0.36$\pm$0.08 and 0.33$\pm$0.07 $M_{\odot}$, and effective temperatures of 3460$\pm$100 and 3410$\pm$100 K, respectively. For USco1621~B and USco1556~B, the combination of luminosity values computed from the direct integration of the spectra with BT-Settl models yields masses and temperatures of 0.015$\pm$0.002 and 0.014$\pm$0.002 $M_{\odot}$, and 2270$\pm$90 and 2240$\pm$100 K for USco1621\,B and USco1556\,B, respectively, assuming an age between 5--10 Myr.

Figure \ref{fig:Lum_vs_age} shows the luminosity vs. age diagram for USco1621\,B and USco1556\,B and other known wide young substellar companions, together with Ames-COND model \citep{Allard2001, Baraffe2003} evolutionary tracks for different masses. Both secondaries are placed slightly above the deuterium-burning mass limit. According to their age, they are entering the deuterium burning phase, where their luminosity will remain with little changes until they reach an age of $\sim$30 Myr. These two new young wide substellar companions are amongst the least massive at separations wider than 1000 AU discovered up to date.

\nocite{SVOVosa}

\section{Discussion}
\subsection{Companionship of the USco1621 and USco1556 systems}

In the previous sections, we  determine that USco1621\,B and USco1556\,B have low-gravity features, which are characteristic of youth. Therefore, both objects belong to the USco star-forming region. However, there is the possibility that both objects belong to USco but are not bound to their primaries and are located at a similar projected position on the sky by chance.
   
To evaluate whether both systems are physically bound or chance alignments, we estimate the probability of finding isolated USco low-mass brown dwarfs at angular separations closer than 25\arcsec of any USco member in our search. Using the survey of similar depth done by \citet{Lodieu2013}, we calculated the density of low-mass brown dwarfs and isolated planets in USco. We find a density of 0.84 obj deg$^{-2}$ for USco members with \textit{J}\textgreater15 mag.

Assuming a poissonian distribution, the probability of finding one or more chance alignments is given by
\begin{equation}
P(x>0)=1-P(x=0)=1-e^{-\lambda}~,
\end{equation}
and the probability of finding two or more chance alignments is given by:
\begin{equation}
P(x>1)=1-P(x=1)-P(x=0)=1-e^{-\lambda}-\lambda \cdot e^{-\lambda},
\end{equation}
with $\lambda=n p$, where $P$ is the poissonian distribution, $x$ is the number of chance alignments found, $n$ is the number of targets in our search (1195) and $p$ is the individual probability of finding a \textit{J}\textgreater15 mag USco member in an area of radius 25\arcsec ($p$=1.27$\times$10$^{-4}$). We obtain a probability of 14\% of finding at least one \textit{J}\textgreater15 mag USco companion candidate by chance alignment. The probability of finding two chance alignments is 1.0\%. 

In conclusion, the probability that both USco1621 and USco1556 systems are bound is 86\%, the probability that at least one of them is bound is 99\%, and the probability that none of them is physically related is 1.0\%. A precise parallax measurement of the secondaries would help confirming their companionship.

\citet{Kraus2008} estimate that for the USco association, true binaries can be distinguished from chance alignments at separations lower than 75\arcsec. Candidate companions closer than this limit, although can also be chance alignments, are most likely bound, and the chance alignment probability is progressively lower as we move to closer separations. The components of USco1621 and USco1556 systems are separated 21\arcsec and 25\arcsec, respectively. Therefore, according to their criteria, these systems are most likely gravitationally bound.

\subsection{Origin and evolution of the systems}

Although several wide substellar companions with separations above 1000 AU have been found around stars, brown dwarfs, and white dwarfs of different ages, the number of such systems is still very limited. An exhaustive compilation of such wide substellar companions is shown in Table \ref{table:widecompilation}.

Several formation scenarios have been proposed to explain the existence of these widely separated systems. It is possible that these very low-mass binaries form in-situ at very wide orbits as a consequence of the fragmentation of the initial molecular cloud, which can result in cores with separations of $\sim$10$^{3}$--10$^{4}$ AU \citep[e.g. ][ and references therein]{Launhardt2010, Chen2013, Goodwin2007}. They could also have formed via disk gravitational instability \citep[e.g. ][]{Boss2001, Whitworth2006}, which can form substellar objects at separations up to several hundreds of AU \citep{Stamatellos2009}, and later migrated to wider separations. In addition, recent observations indicate that a large fraction of wide, low-mass binaries are in fact triple systems, where the more luminous component is a binary itself \citep[e.g.][]{Elliott2016}. One possible explanation of this behaviour is that the three components may form in a compact system, and then one of them, typically the least massive, evolves to wider orbits by three-body dynamical interactions \citep{DelgadoDonate2004, Reipurth2012}. Another possible mechanism for the existence of very wide binaries is the formation of both components in separate locations and their later gravitational capture \citep{Kouwenhoven2010}. It is not clear which of these mechanisms may be the origin of the USco1621 and USco1556 systems.

To understand the possible future evolution of these systems, we calculated their binding energies and estimated their disruption times. For the binding energies we used the relation

\begin{equation}
U = \frac{-G \, m_{1} \, m_{2}}{a},
\end{equation}
where $G$ is the gravitational constant, $m_{1}$, $m_{2}$ are the masses of the binary components  and $a$ is the semi-major axis of the orbit. It is important to note that, due to projection effects, the real semi-major axis of the orbit may be larger than the separation we measure here. \citet{Fischer1992} made a probabilistic calculation of the average value of the semi-major axis, by performing a Monte Carlo simulation which included all the possible orbital parameters for any elliptical orbits in randomly oriented inclinations. They obtain an average semi-major axis of $\langle a_{rel} \rangle = 1.26 d \langle \alpha \rangle$, where $d$ is the heliocentric distance and $\langle \alpha \rangle$ is the average observed angular separation of the components. Hence, for this average semimajor axis we obtain a binding energy of $(-2.6 \pm 1.0) \times 10^{33} \, $J and $(-1.8 \pm 0.7) \times 10^{33} \, $J for USco1621 and USco1556 systems, respectively, and an (absolute) upper limit of $(-3.3 \pm 1.2) \times 10^{33} \, $J and $(-2.3 \pm 0.9) \times 10^{33} \, $J for USco1621 and USco1556 systems, respectively, using the observed separation. This low binding energy is comparable to Oph 16222$-$2405 \citep[$-$1.6 $\times$ 10$^{33}$ J;][]{Close2007}, UScoCTIO 108 AB \citep[$-$1.9 $\times$ 10$^{33}$ J;][]{Bejar2008}, or the ultrawide systems 2MASS J0126AB \citep[$-$3.0 $\times$ 10$^{33}$ J;][]{Artigau2007, Caballero2009} and 2M1258+40 \citep[$-$2.5 $\times$ 10$^{33}$ J;][]{Radigan2009}. 

To estimate the disruption times due to dynamical encounters, we used the soft binary approximation from \citet{BinneyTremaine}, given by
\begin{equation}
t_{\mathrm{evap}} = \frac{m_{1} + m_{2}}{m_{a}} \frac{\sigma}{16 \sqrt{\pi} \, G \, \rho \, a \, ln \Lambda},
\end{equation}
where $m_{1}$, $m_{2}$ are the masses of the binary components, $m_{a}$ is the mean mass of the perturbers, $\sigma$ is the velocity dispersion, $G$ is the gravitational constant, $\rho$ is the mass density of the environment, $a$ is the semimajor axis of the orbit of the binary, and $ln \, \Lambda$ is the Coulomb logarithm, which can be calculated using the expression

\begin{equation}
\Lambda = \frac{b_{\mathrm{max}} V_{0}^{2}}{G (M+m)},
\end{equation}
with $b_{\mathrm{max}}$ being the maximum impact parameter that needs to be considered in the perturbations, $V_{0}^{2}$ the relative velocity of the perturber, $M$ the mass of the object (total mass of the binary in our case) and $m$ the mass of the perturber. For $b_{\mathrm{max}}$ we used the expression

\begin{equation}
b_\mathrm{max}(a) = 9.6 ~\mathrm{pc} \left( \frac{v_{\mathrm{rel}}}{20 ~\mathrm{km} \cdot \mathrm{s}^{-1}} \right) \left( \frac{a}{0.1 ~\mathrm{pc}} \right)^{3/2} \left( \frac{M}{M_{\odot}} \right)^{-1/2}
\end{equation} 
\citep{Weinberg1987}, where $v_{\mathrm{rel}}$ is the relative velocity between the object and the perturber, $a$ is the separation between the binary components, and $M$ is the total mass of the binary.

In addition, we also estimated the disruption time using the approximation in \citet{Weinberg1987}, given by

\begin{equation}
\begin{aligned}
t_{*} & = 1.8 \cdot 10^{4} ~\mathrm{Myr} \left( \frac{n_{*}}{0.05 ~\mathrm{pc}^{-3}} \right)^{-1} \left( \frac{M}{M_{\odot}} \right) \left( \frac{M_{*}}{M_{\odot}} \right)^{-2} \cdot \\
& \cdot \, \left( \frac{<1/v_{\mathrm{rel}}>^{-1}}{20 ~\mathrm{km} \cdot \mathrm{s}^{-1}} \right) \left( \frac{a_{0}}{0.1 ~\mathrm{pc}} \right)^{-1} \ln^{-1} {\Lambda},
\end{aligned}
\end{equation}
where $n_{*}$ is the density of perturbers, $M$ is the total mass of the binary, $M_{*}$ is the mean mass of the perturbers, $v_{\mathrm{rel}}$ is the relative velocity between the binary and the perturbers, $a_{0}$ is the separation of the binary, and $ln \, \Lambda$ is the Coulomb logarithm.
We considered a mean perturber mass of $1 M_{\odot}$ and a mean velocity dispersion of 3.20  km s$^{-1}$ \citep{Wright2018}. 
\citet{Wright2018} did not find any clear evidence of expansion neither in USco nor in any of the subgroups of Sco-Cen association. This means that it is probable that Sco-Cen was formed during several formation events and not as a single dense cluster that later expanded into a larger region. In this case, we can consider that the actual estimated density in their environments has not significantly changed since their births. To estimate the current density of objects around each of the systems, we used the astrometric and parallactic
data from $Gaia$ DR2 to select all the targets (belonging to USco or not) located inside a cube of 8 pc side ($\sim$500 pc$^{3}$) around each system. We found 259 and 172 objects within these volumes, which correspond to densities of 0.51 and 0.34 obj pc$^{-3}$ around the USco1621 and USco1556 systems, respectively. For the binary separation, we considered the average semimajor axis, $\langle a_{rel} \rangle = 1.26 d \langle \alpha \rangle$, and the measured separation as a lower limit. For these values, the estimated disruption times range between 90--250 and 100--260 Myr, respectively. Therefore, if these systems are really physically bound, they are expected to survive for several Myr in their actual environments. However, at these timescales, the USco association is expected to already be dissipated. In that case, considering a density similar to that of the Solar vicinity in their future environments \citep[0.076--0.084 systems pc$^{-3}$, calculated using data from The Research Consortium On Nearby Stars, RECONS, in][]{Henry2018}, the systems are expected to survive for $\sim$0.4--1.7 Gyr.

We have also investigated the possibility that these systems have already suffered a close encounter and are currently in the process of disruption. Using $Gaia$ DR2, we have performed a search for nearby objects which could have interacted with the systems in the last thousands of years. Given the minimum escape velocity of the companions at the present location ($\sim$0.4 km s$^{-1}$), if they are currently in disruption, the perturbation must have happened less than $\sim$0.1 Myr ago because, otherwise, the companions would have traveled a distance greater than their actual separation. We identified all the nearby objects with heliocentric distances corresponding with those of the primaries $\pm$5 pc, and using their proper motions we traced back their 2-D trajectories over the last 0.3 Myr. The $\pm$5 pc limit implies that we are considering all the possible perturbers (from USco or the field population) with radial velocities differing up to $\sim\pm$50 km s$^{-1}$ with respect to the systemic velocity of our systems, because this is the maximum distance the perturber would have travelled in 0.1 Myr. We find several candidates that may have been located within 1~pc of the primaries in the past, but the lack of radial velocity information prevents us from confirming if they were really physically close to the systems in a 3D picture. As a result of this search, we have identified a promising perturber candidate for the USco1621 system. This object is a candidate member of the USco association, 2MASS J16223009$-$2532319 \citep{Luhman2018, Damiani2019}, whose distance, 139$\pm$2 pc, is compatible with USco1621~A. Their projected separation is 0.88 pc (182\,000 AU), and their proper motions are currently diverging, with a maximum approximation just three years ago. The apparent magnitudes and colours of 2MASS J16223009$-$2532319 indicate an expected spectral type of mid-M. Given the large physical separation of this object to the binary system USco1621~AB, it is difficult to determine if it is part of a triple system or just an USco member located by chance at this separation.
    
\subsection{Adding two new benchmark systems}   
   
In USco, several tens of substellar objects with masses around or below the deuterium burning limit mass have been found in isolation \citep{Lodieu2006, Lodieu2007, Dawson2011, Lodieu2013, PenaRamirez2016, Lodieu2018}. However, only four of these objects have been found as wide companions: UScoCTIO 108 B \citep{Bejar2008}; 1RXS1609b \citep{Lafreniere2008}; GSC 06214-00210 \citep{Ireland2011}; and HIP 77900B \citep{Aller2013}. Two of them (1RXS1609b and GSC 06214-00210) were discovered using adaptive optics and their low angular separations from their brighter primaries make it difficult to fully characterise them, especially at optical wavelengths, allowing only limited spectral characterisation in the infrared. The discovery of USco1621~B and USco1556~B has added two new benchmark objects to this small collection, which are easier to characterise, also in the optical wavelength range.

Other companions in wide orbits have been identified populating distinct young associations and stellar moving groups of similar ages. Some examples are FU Tau B \citep{Luhman2009} and SCH06 J0359+2009 B \citep{Kraus2012, Martinez2019} in Taurus; SR12 C \citep{Kuzuhara2011} in Ophiucus; HD 106906 b \citep{Bailey2014} in Lower Centaurus Crux; or AB Pic B \citep{Chauvin2005} and 2MASS J2126$-$8140 \citep{Deacon2016} in Tucana-Horologium.


\section{Summary and final remarks}

The main results of this paper are summarised below:

   \begin{enumerate}
      \item Using VHS and UKIDSS GCS data, we  found two new wide substellar companions (USco1621\,B and USco1556\,B)  of two previously-known early-M dwarfs in USco. The companions are located at angular separations of 21\arcsec and 25\arcsec\  which, at the distance of their primaries, correspond to projected physical separations of 2880 and 3500 AU, respectively.
      \item Using low-resolution spectroscopy, we classified USco1621\,B and USco1556\,B as M8.5$\pm$0.5 in the optical and L0$\pm$0.5  and L0.5$\pm$0.5  in the near-infrared, respectively.
       \item We find later spectral types in the near-infrared than in the optical by 1.5 to 2.0 subclasses. Similar differences have been reported by earlier works on this association.
      \item These objects show features of youth in their optical and infrared spectra compatible with the age of USco, which confirms their very young age and their membership to this association.
      \item Using high-resolution images from the Keck-I LGS adaptive optics system, we did not find any additional stellar or brown dwarf companion at separations larger than 0.3" ($\sim$40 AU) from the primaries and any additional planetary-mass companion at separations larger than 0.2" from the secondaries ($\sim$28 AU).
            \item We derived the bolometric luminosities of USco1621\,B and USco1556\,B by integrating their spectra and using the distances of the primaries, measured by $Gaia$ DR2. We obtain luminosities of $\log{(L/L_{\odot})}$=$-$3.03$\pm$0.11 and $-$3.07$\pm$0.11 for USco1621\,B and USco1556\,B, respectively. According to theoretical evolutionary models, we obtain masses of 0.015$\pm$0.002 and 0.014$\pm$0.002 $M_{\odot}$, and temperatures of 2270$\pm$90 and 2240$\pm$100~K, for USco1621\,B and USco1556\,B, respectively. These estimations confirm that they are low-mass brown dwarfs, with masses slightly above the deuterium burning mass limit.
       \item We estimated that the probability that both systems are bound systems belonging to USco rather than chance alignments is 86\%, the probability that at least one of them is physically bound is 99\%, and the probability that none of them are bound is 1.0\%. 
       \item Considering the current densities in the surroundings of each system, we estimated disruption times of few hundreds of Myr for USco1621 and USco1556 systems, which is larger than the current age of USco and its expected dissipation time. Considering a density similar to that of the Solar vicinity after the dissipation of the USco association, the systems are expected to survive for $\sim$0.4--1.7 Gyr.

   \end{enumerate}

\begin{acknowledgements}
      We thank the anonymous referee for his/her very useful corrections and suggestions, which helped improving this manuscript.
      P.C., V.J.S.B. and N.L. are partially supported by grant AyA2015-69350-C3-2-P; R.R. by program AyA2014-56359-P; M.R.Z.O. by program AyA2016-79425-C3-2-P and A.P.G by program AyA2015-69350-C3-3-P from the Spanish Ministry of Economy and Competitiveness (MINECO/FEDER). B.G. acknowledges support from the CONICYT through FONDECYT Postdoctoral Fellowship grant No 3170513.
      This paper is based on observations performed at the European Southern Observatory (ESO) in Chile, under programs 092.C-0874 (PI Gauza), 099.C-0848 (PI Chinchilla) and 0101.C-0389 (PI Chinchilla). This work is based on observations (program 55-GTC30/17A; PI Lodieu) made with the Gran Telescopio Canarias (GTC), operated on the island of La Palma in the Spanish Observatorio del Roque de los Muchachos of the Instituto de Astrof\'isica de Canarias. Some of the data presented herein were obtained at the W.M. Keck Observatory, which is operated as a scientific partnership among the California Institute of Technology, the University of California and the National Aeronautics and Space Administration. The Observatory was made possible by the generous financial support of the W.M. Keck Foundation. The authors wish to recognise and acknowledge the very significant cultural role and reverence that the summit of Maunakea has always had within the indigenous Hawaiian community.  We are most fortunate to have the opportunity to conduct observations from this mountain.
      This research has made use of the Simbad and Vizier \citep{SIMBADpaper} data bases, operated at the Centre de Donne\'es Astronomiques de Strasbourg (CDS), and NASA's Astrophysics Data System Bibliographic Services (ADS). This work has made use of data from the European Space Agency (ESA) mission Gaia (\url{https://www.cosmos.esa.int/gaia}), processed by the Gaia Data Processing and Analysis Consortium (DPAC, \url{https://www.cosmos.esa.int/web/gaia/dpac/consortium}). Funding for the DPAC has been provided by national institutions, in particular the institutions participating in the Gaia Multilateral Agreement. This research has made use of the SVO Filter Profile Service (\url{http://svo2.cab.inta-csic.es/theory/fps/}) supported from the Spanish MINECO through grant AYA2017-84089. This publication made use of Python programming language (Python Software Foundation, \url{https://www.python.org}).

\end{acknowledgements}







\bibliographystyle{aa} 
\bibliography{biblio.bib} 

\end{document}